# Data-driven local operator finding for reduced-order modelling of plasma systems: I. Concept and verifications

F. Faraji*[1], M. Reza*, A. Knoll*, J. N. Kutz**

* Plasma Propulsion Laboratory, Department of Aeronautics, Imperial College London, London, United Kingdom

** Department of Applied Mathematics and Electrical and Computer Engineering, University of Washington, Seattle, United States

**Abstract**: Computationally efficient reduced-order plasma models able to reliably predict the plasmas behavior across different settings and configurations in a self-consistent manner have remained unachievable so far. The necessity for these models has nonetheless continuously increased over the past decade. This is because the reduced-order models (ROMs) can, on the one hand, facilitate scientific research into yet-unresolved complex plasma phenomena. On the other hand, predictive ROMs can streamline and expedite the development of advanced, high-efficiency plasma technologies. With the increase in computational power in recent years, and the emergence of several approaches that lower the computational burden of generating extensive high-fidelity plasma datasets, data-driven dynamics discovery methods can play a transformative role toward the realization of predictive, generalizable, and interpretable ROMs for plasma systems. In this two-part article, we introduce a novel data-driven algorithm – "Phi Method" – for the discovery of discretized systems of differential equations describing the dynamics. The success and generalizability of Phi Method roots in its constrained regression on a library of candidate terms that is informed by numerical discretization schemes. Phi Method's performance toward the derivation of reliable and robust ROMs are demonstrated in this part I for three test cases – the Lorenz attractor problem, the flow-past-a-cylinder problem, and a 1D Hall-thruster-representative plasma problem. Part II will focus on Phi Method's application for parametric dynamics discovery. The performance of ROMs from Phi Method is compared for reference against the performance of ROMs from Optimized Dynamic Mode Decomposition (OPT-DMD) method in the fluid and the plasma test cases. For all the test cases, Phi-Method-derived ROMs provided remarkably accurate predictions of the systems' behavior in terms of the evolution dynamics of the involved state variables. It was further demonstrated that accurate ROMs can be reliably developed by applying Phi Method to either steady-state or transient-state data from a system.

**Section 1: Introduction**

In contemporary scientific research and industrial developments surrounding the plasma systems, there is a growing recognition of the necessity for reliable, self-consistent, and computationally efficient reduced-order plasma models. On the one hand, the reduced-order models are more interpretable than the high-dimensional kinetic plasma simulations, for instance, that feature significant degrees of freedom. These models can, thus, facilitate our endeavors to fully understand the underlying physics of plasmas and to address the long-standing questions in the field, such as the evolution of the instabilities and their interactions as well as their impacts on the plasma species transport across the magnetic fields [1]-[3]. On the other hand, reliable and cost-effective reduced-order models (ROMs) will majorly enhance our ability to predict and control the plasmas, offering potentially groundbreaking advancements in today's plasma technologies as well as toward the exploration of new applications across various sectors for the future.

Despite the great interest in and demand for reliable predictive and generalizable ROMs for plasmas, these models have remained elusive so far. Within the plasma modelling community, a mainstream approach toward such models has been one that entails the derivation of first-principles models [4][5], which numerically solve the conservation equations derived from the moments of the plasma kinetic equation [6]. These so-called "fluid-based" models have a low computational cost compared to the kinetic class of plasma simulations, which are associated with significant computational cost and high-dimensionality [7][8]. The fluid models can be thought of as reduced-order models in the sense that they reduce the dimensionality of the problem solved by a kinetic simulation, e.g., 3D3V in the phase space, to 3D in the configuration space through taking moments of the plasma kinetic equation. Despite their low computational cost and reduced-dimensionality, the first-principles fluid-based models still lack universally applicable physics-based closures [7][9][10][11] and are also unable to capture kinetic processes like the microscopic instabilities and their induced momentum and energy transport [7]. As a result, these plasma models remain reliant on case-specific ad-hoc relations tuned on experimental data to close the involved system of equations and/or to incorporate the effects of the unresolved physics [4][7].

---

[1] **Corresponding Author** (f.faraji20@imperial.ac.uk)

Recently, a different path toward predictive, interpretable, and generalizable ROMs has also been pursued by plasma physics researchers. This path consists of using machine-learning (ML)/data-driven (DD) algorithms applied to the datasets from high-fidelity simulations [12]-[17]. The enabling elements of this approach are: (1) cost-efficient high-fidelity simulations that enable the generation of representatively large datasets for the training of ML/DD algorithms, (2) robust and reliable ML/DD architectures.

Regarding element 1, one of the highest-fidelity plasma modelling approaches is the particle-in-cell (PIC), which is a statistical approach to kinetically resolve the evolution of the plasma species [18][19]. The PIC simulations are of sufficient fidelity to self-consistently capture all relevant phenomena within the plasma across the involved scales. Thus, high-fidelity data generated from the PIC simulations can provide valuable information to derive closure relations and ROMs since these data represent the complex interdependencies between different plasma variables and the interactions among the underlying plasma processes. Nonetheless, the enormous computational cost of the conventional PIC simulations renders them impractical for the generation of sufficiently broad sets of multi-dimensional data that span over the large spatial and temporal scales typically present within real-world plasma systems.

In any case, in recent years, several efforts have aimed at tackling the computational cost issue of the kinetic particle-in-cell simulations. These efforts include the development of implicit and/or energy-conserving PIC algorithms [20][21], GPU implementation of the PIC method [22], the sparse-grid PIC scheme [23], and the reduced-order PIC scheme [24]. The last approach, developed at Imperial Plasma Propulsion Laboratory (IPPL) and extensively verified across a variety of test cases (plasma configurations) [25]-[28], reduces the computational cost of conventional PIC simulations by up to 3 orders of magnitude and, hence, paves the path to highly cost-efficient generation of high-fidelity training datasets for ML/DD algorithms.

Considering element 2, i.e., robust and reliable ML/DD architectures, data-driven dynamics discovery approaches have catalyzed a paradigm shift over the past years in understanding and uncovering the inherent dynamics of various systems, on the one hand, and toward the development of predictive reduced-order models, on the other, by leveraging the information embedded within data from the system itself. The DD dynamics discovery techniques can be generally divided into two categories [29][30]: one in which the methods find the global structures and the dynamics across the entire system, and the other one in which the algorithms focus on the identification of local operators, thereby discovering the governing system of differential equations.

In the first category, Dynamic Mode Decomposition (DMD) [31][32] is proven to be a valuable tool in fluid dynamics [33][34] and, more recently, in plasma physics [14][35]-[37]. DMD enables the identification of dominant coherent structures/modes in a given flow field of fluid/plasma systems [14][31][38]. The DMD's utility for the development of ROMs capable of forecasting the state of dynamical systems is also demonstrated [39][40]. The OPT-DMD method [41], which is designed to overcome the inherent sensitivity of basic DMD to noise in the data, is further shown to yield a significant improvement in the robustness and the predictive capabilities of the derived DMD ROMs. In this regard, in Refs. [12][13], we demonstrated the promising predictive potential of the ROMs from OPT-DMD in a 2D plasma configuration representative of a radial-azimuthal cross-section of a Hall thruster discharge. Across a range of plasma conditions in the adopted 2D radial-azimuthal configuration, the linear OPT-DMD-derived ROMs showed a remarkable ability to reconstruct and predict the ground-truth data from the PIC simulations, especially in cases where the plasma system exhibited a relatively quasi-periodic dynamics [12][13].

Regarding the second category, two prominent examples of the well-established approaches that enable finding the nonlinear relations and the differential equations describing the spatiotemporal evolution of the systems include: (i) symbolic regressions using Genetic Optimization [42][43], and (ii) Sparse Identification of Nonlinear Dynamics (SINDy) [44]-[46]. It is noted that, contrary to the computationally expensive symbolic regression using genetic programming, SINDy offers a more cost-effective alternative via sparse regression on a predefined library of candidate terms, which can potentially describe a system's dynamics [44].

In this two-part article, we introduce and demonstrate a data-driven regression-based local-operator discovery algorithm, that is developed at IPPL and that we have termed "Phi Method". Phi Method enables learning the discretized form of the partial differential equations describing a dynamical system and, thus, allows straightforward time forecasting of the system's state. The data-driven finding of the optimum discretization stencil embedded in Phi Method is inspired from the general "stencil-learning" research in fluid dynamics [47][48], where ML/DD approaches have been pursued to relax the stringent stability requirements on solving the differential equations with nonlinear terms (representing turbulence e.g.) that make the numerical solutions computationally burdensome.

In this part I, we first present the concept underpinning the Phi Method algorithm and then verify the algorithm's generalizability and performance for ROM development in multiple test cases, namely, the Lorenz attractor, the 2D problem of a fluid flow past a cylinder, and a plasma problem corresponding to the 1D azimuthal dynamics of a Hall thruster discharge.

In the Lorenz attractor test case, our aim is to show the Phi Method's capability to recover dynamics from trajectory data that are governed by a system of nonlinear ordinary differential equations (ODEs). The remaining two test cases involve a set of partial differential equations (PDEs). The fluid system serves as a benchmarking case, in which we aim to demonstrate the performance of the Phi-Method-derived ROM toward forecasting the fluid flow field in comparison against an OPT-DMD-derived ROM. In the last test case, the goal is to demonstrate the applicability of the ROMs from Phi Method for plasma systems and toward simultaneously forecasting the plasma state variables. An OPT-DMD model is also presented for the plasma discharge test case for two purposes: first, as a comparison reference for the Phi Method model, and second, to allow us to further assess the effectiveness of OPT-DMD for ROM development in plasma applications by extending the studies in Refs. [12][13] to a new plasma test case.

**Section 2: Description of the data-driven methods**

Before introducing the Phi Method's concept, we present an overview of the DMD approach in Section 2.1. This overview is particularly important because it allows us to clarify the relationship between Phi Method and DMD later in Section 2.2, where we discuss the idea and the formulation that underpin Phi Method.

**2.1. Dynamic Mode Decomposition**

Dynamic Mode Decomposition is fundamentally a linear-operator based technique to decompose spatiotemporal datasets into vector bases that capture coherent spatial structures and temporal dynamics in the data [31][32]. At its core, DMD leverages the Singular Value Decomposition (SVD) to reduce the dimension of the high-dimensional datasets and to extract the dominant spatial features. Moreover, DMD operates under the assumption of linear dynamics. The linearity constraint allows the spatiotemporal modes to be uniquely determined. As a result, DMD can provide a best-fit linear-time-dynamics model describing the spatiotemporal data from high-fidelity simulations or experiments. It provides low-rank and interpretable ROMs in the form of dominant modal structures along with their exponential/oscillatory (linear) behavior.

DMD can be applied to the spatiotemporal signals of each individual state variable of a system separately. However, DMD is also able to work with multi-property state vectors. In plasma applications, for example, assembling a multi-property plasma state comprised by several plasma properties can amount to representing the full state vector of the plasma system. This will then allow us to develop a linear ROM that best describes the time evolution of all the properties (state variables) simultaneously. The enhancement of the state vector in this manner enables capturing the interdependency between the plasma properties and is, thus, expected to improve the predictive performance of the derived ROMs.

The DMD algorithm starts by constructing matrices of time-series data snapshots $X$ and $X'$ as per in Eq.1.

$$X = \begin{bmatrix} | & | & & | & & | \\ \mathcal{X}_1 & \mathcal{X}_2 & \ldots & \mathcal{X}_k & \cdots & \mathcal{X}_{m-1} \\ | & | & & | & & | \end{bmatrix}, \quad X' = \begin{bmatrix} | & | & & | & & | \\ \mathcal{X}_2 & \mathcal{X}_3 & \ldots & \mathcal{X}_{k+1} & \cdots & \mathcal{X}_m \\ | & | & & | & & | \end{bmatrix}. \quad \text{(Eq. 1)}$$

Referring to Eq. 2, each column $\mathcal{X}_k$ in $X$ represents a column-wise stacking of the rearranged snapshots of several plasma properties at time $t_k$ ($x_{p,k}$), with $p$ denoting the index of a single specific property, where $p = 1,2,\ldots,P$. $x_{p,k}$ is a snapshot of p-th property at time $t_k$ ($k = 1,2,\ldots m-1$) rearranged into a $n$-dimensional column vector. Also, $m$ indicates the number of timesteps over which the data is collected from a system.

$$\mathcal{X}_k = \begin{bmatrix} x_{1,k}^T, & x_{2,k}^T, \ldots, x_{P,k}^T \end{bmatrix}^T. \quad \text{(Eq. 2)}$$

The corresponding column $\mathcal{X}_{k+1}$ in $X'$ is the system's state vector at the subsequent time step $t_{k+1}$.

The DMD then approximates the optimal (in the least square sense) linear operator $A$ that simultaneously advances all data snapshots one time-step forward such than

$$X' \approx AX, \tag{Eq. 3}$$

and,

$$A = \arg\min_{A} \|X' - AX\|_2 \approx X'X^{\dagger}. \tag{Eq. 4}$$

In Eq. 4, $X^{\dagger}$ is Moore-Penrose pseudo-inverse of $X$, which can be obtained using the Singular Value Decomposition (SVD) of matrix $X$ as in Eq. 5.

$$X = U\Sigma V^*. \tag{Eq. 5}$$

In Eq. 5, $U \in C^{n \times m}$, and its columns are the left singular vectors ($u_i$) of matrix $X$ representing the SVD/POD modes. $V \in C^{m \times m}$, whose columns represent the right singular vectors ($v_i$) of matrix $X$. $V^*$ indicates conjugate transpose of matrix $V$. Also, $\Sigma \in R^{m \times m}$ is a diagonal matrix with non-negative entries along its diagonal and holds the singular values ($\sigma_i$) associated with the SVD/POD modes.

In practice, matrix $X$ can be approximated by a low-rank truncation of the SVD series. The rank-$r$ truncated approximation of matrix $X$ is achieved by retaining the first $r$ terms in the SVD expansion as in the following

$$\tilde{X}_r = \sum_{i=1}^{r} \sigma_i u_i v_i^* = \tilde{U}\tilde{\Sigma}\tilde{V}^*, \tag{Eq. 6}$$

where, $\tilde{U} \in C^{n \times r}$, $\tilde{\Sigma} \in R^{r \times r}$ and $\tilde{V} \in C^{m \times r}$. Similarly, it is more efficient for DMD to compute the low-rank approximation of operator $A$, which is denoted as $\tilde{A}$. As a result, the $r \times r$ projection of full matrix $A$ onto the first $r$ SVD modes of matrix $X$ can be obtained as

$$\tilde{A} = \tilde{U}^* A \tilde{U} = \tilde{U}^* X' \tilde{V} \tilde{\Sigma}^{-1}. \tag{Eq. 7}$$

The eigenvalues and the eigenvectors of $\tilde{A}$ will be determined following the eigendecomposition of the matrix $\tilde{A}$

$$\tilde{A}V = \Lambda V, \tag{Eq. 8}$$

where, $\Lambda$ is a diagonal matrix containing the eigenvalues ($\lambda_i$, i = 1, 2, ..., $r$) of $\tilde{A}$. In addition, the columns of matrix $V$ are the eigenvectors of $\tilde{A}$, which can be expanded to reconstruct the eigenvectors of $A$. The eigenvectors of $A$ are represented by the columns of the matrix $\Psi$ ($\psi_i$, i = 1, 2, ..., $r$), which is obtained using Eq. 9.

$$\Psi = X'V\Sigma^{-1}V \tag{Eq. 9}$$

The eigenvectors $\psi_i$ represent the DMD spatial modes and their corresponding eigenvalues $\lambda_i$ characterize the modes' temporal dynamics. Once the DMD spatiotemporal modes are determined, we can approximate the time-series data using the DMD expansion through the relation in Eq. 10.

$$x_{k+1} \approx \sum_{i=1}^{r} b_i \psi_i \lambda_i^k = \Psi B \Lambda^k, \qquad k = 1, 2, \dots. \tag{Eq. 10}$$

In Eq. 10, matrix B is a diagonal matrix that contains the initial amplitudes of the DMD modes ($b_i$ is the initial amplitude associated with i-th DMD mode, $\psi_i$). B is obtained by projecting the data at time $t = 0 - x_1$ – onto the DMD modes, i.e.,

$$B = \Psi^{\dagger} x_1. \tag{Eq. 11}$$

One of the major limitations of the basic DMD approach, as was described above, is its significant sensitivity to noise in the data. Considering that the real-world data, either from experimental measurements or from statistical particle-based simulations, inherently contain noise, the induced noise biases can hinder the effective application of DMD for the development of reliable ROMs [41]. The susceptibility of DMD to noise stems from its exclusive reliance on pairwise relationships between consecutive state vectors ($\mathcal{X}_k$ and $\mathcal{X}_{k+1}$). The presence of noise can distort these relationships which can then mislead the derivation of the spatiotemporal modes.

In view of this shortcoming, OPT-DMD method [41] provides a remarkable advancement over the basic DMD. To alleviate the sensitivity of the DMD approach to noise, OPT-DMD considers the collective relationship among all snapshots simultaneously [41]. The OPT-DMD algorithm solves the nonlinear optimization problem given by Eq. 12 using the variable projection technique [41] to find the optimum spatial modes ($\Psi_B = \Psi B$) and the temporal dynamics ($\Omega$, italic omega) simultaneously.

$$\min_{\Omega, \Psi_B} \|X - \Psi_B \exp(\Omega t)\|_2 \tag{Eq. 12}$$

$\Omega$ represents the DMD modes' dynamics (frequencies, $\omega_i$) in the continuous-time domain, which can be translated to the discrete-time dynamics as represented by entries of the matrix $\Lambda$ ($\lambda_i$), using the relation $\lambda_i = \exp(\omega_i \Delta t)$.

Another limitation of the basic DMD algorithm is its underlying assumption of the linearity of the time dynamics (Chapter 10 of Ref. [32]). Nonetheless, nonlinearities can be introduced into a DMD model by expanding the state vector's space so that it includes nonlinear observables (functions) of the state variables as well. To this end, the state vectors, $\mathcal{X}_k$s, in the data matrices $X$ and $X'$ can be augmented to include the state variables themselves as well as their nonlinear observables, $g_i$, $i = 1, 2, \ldots, N$ (Eq. 13) [32].

$$\mathcal{X}_k = \left[x_{1,k}^T, \ x_{2,k}^T, \ldots, x_{P,k}^T, \ g_1(x_k^T), \ g_2(x_k^T), \ldots, \ g_N(x_k^T)\right]^T, \tag{Eq. 13}$$

where,

$$\pmb{x}_k = \left[x_{1,k}^T, \ x_{2,k}^T, \ldots, x_{P,k}^T\right]. \tag{Eq. 14}$$

The above approach is inspired by the Koopman theory, which refers to representing a *finite-dimensional* nonlinear dynamical system using an *infinite-dimensional* linear transformation known as the Koopman operator [49]. The Koopman operator acts on an infinite-dimensional state space that includes all the possible nonlinear observables of the physical state variables of the system.

The concepts of an infinite-dimensional operator and state space are, however, an abstraction. In practice, we need to find a *finite-dimensional approximation* of the Koopman operator by choosing a finite observable subspace. Choosing an appropriate subset of the observables, which enables to best approximate the full Koopman transformation, may require expert knowledge. In the absence of an a-priori knowledge, techniques such as Kernel-based methods and, in particular, Support Vector Machines (SVM) [50]-[53] can be used to find the optimal subset of observables.

### 2.2. Phi Method

Phi Method is a novel data-driven dynamics discovery technique, which finds the discretized form of PDEs governing a system's dynamics. In other words, the algorithm is designed to discover the involved dynamics and the discretization stencil simultaneously. This, hence, yields an optimum local operator that acts as a time stepper, advancing the system's state vector through time. The underlying formulation of Phi Method is presented in the following. For simplicity, and without losing generality, we describe the formulation for the case of a system with one-dimensional (1D) spatial variations.

Consider the relation in Eq. 15, which describes the dynamics (time evolution) of a system

$$\frac{d\pmb{x}(t)}{dt} = \pmb{\mathcal{F}}\big(\pmb{x}(t)\big), \tag{Eq. 15}$$

with $\pmb{x}$ denoting the system's state vector. In the context of plasma applications, e.g., $\pmb{x}$ can represent all the plasma properties of interest from the system. We aim to discover the discretized form (in time and space) of Eq. 15, which can be written as follows

$$\pmb{x}_i^{k+1} = \pmb{F}(\pmb{y}_i^k)\Phi. \tag{Eq. 16}$$

In Eq. 16, $\Phi$ represents the linear discrete-form counterpart of the continuous-time operator $\pmb{\mathcal{F}}$ (Eq. 15), which we would like to find, hence, the name of the algorithm, "Phi Method". The relationship in Eq. 16 is applicable to all the nodes, $i$, inside a computational domain ($i = 2, 3, \ldots, N_g - 1$, with $N_g$ being the total number of computational nodes), and all the time steps $k = 1, 2, \ldots, m - 1$. Referring to Eq. 17, $\pmb{x}_i^k$ is the system's state vector comprised by $P$ variables ($x_{p,i}^k$, $p = 1,2, \ldots, P$) on the $i$-th node and at the time step $k$. $\pmb{y}_i^k$ represents the set of state vectors on node $i$ and on its neighboring nodes at the time step $k$.

$$\boldsymbol{x}_i^k = \begin{bmatrix} x_{1,i}^k \\ x_{2,i}^k \\ \vdots \\ x_{P,i}^k \end{bmatrix}^T, \qquad \boldsymbol{y}_i^k = [\boldsymbol{x}_{i-1}^k \ \ \boldsymbol{x}_i^k \ \ \boldsymbol{x}_{i+1}^k]. \tag{Eq. 17}$$

Here, in the definition of $\boldsymbol{y}$, we have included only the immediate adjacent nodes to node $i$ along a single dimension, i.e., nodes $i-1$ and $i+1$. However, the definition can be readily generalized to contain arbitrary number of surrounding nodes along all the dimensions possibly involved in the system's dynamics. The inclusion of farther neighboring nodes (e.g., nodes $i-2, i+2$, etc) allows capturing higher-order spatial derivatives (if present) as well as to obtain higher accuracy discretization of the lower-order derivatives. Similarly, the $\boldsymbol{y}$ vector can also be generalized to include the set of state vectors at the time steps $k-1, k-2$, etc. The schematic of the spatiotemporal computational stencil in a 1D configuration is illustrated in Figure 1.

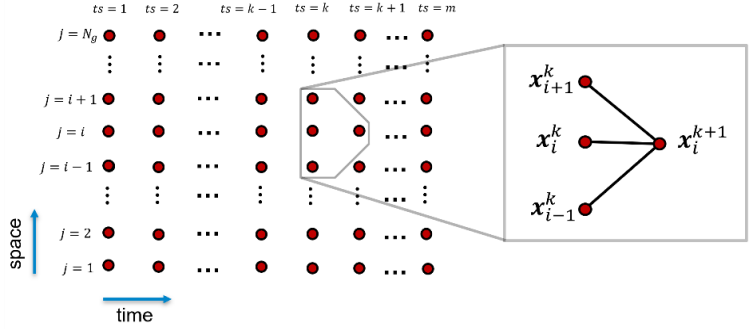

Figure 1: Schematic of the 1D spatiotemporal stencil of the computational nodes. $j$ denotes the node number, and $t_s$ denotes the timestep.

In Eq. 16, $\boldsymbol{F}$ represents a library of candidate terms that includes the linear and nonlinear "observable" functions of $\boldsymbol{y}$, and is defined as below

$$\boldsymbol{F}(\boldsymbol{y}_i) = \begin{bmatrix} f_1(\boldsymbol{y}_i) \\ f_2(\boldsymbol{y}_i) \\ \vdots \\ f_L(\boldsymbol{y}_i) \end{bmatrix}^T, \qquad \boldsymbol{f}_l(\boldsymbol{y}_i) = \begin{bmatrix} f_l(\boldsymbol{x}_{i-1}) \\ f_l(\boldsymbol{x}_i) \\ f_l(\boldsymbol{x}_{i+1}) \end{bmatrix}^T \ ; \ l = 1, 2, \ldots L, \tag{Eq. 18}$$

where $L$ denotes the total number of distinct observable functions ($\boldsymbol{f}$) in the library.

To determine $\Phi$ from a system's data, i.e., in the training phase, we construct data matrices $X$ and $Y$ (Eqs. 19 and 20), which, respectively, contain the state vectors $\boldsymbol{x}$ and $\boldsymbol{y}$ evaluated on different nodes and over several time steps

$$X = \left[\boldsymbol{x}_2^2, \boldsymbol{x}_3^2, \ldots, \boldsymbol{x}_{N_g-1}^2, \boldsymbol{x}_2^3, \boldsymbol{x}_3^3, \ldots, \boldsymbol{x}_{N_g-1}^3, \ldots \ldots \ldots \ldots, \boldsymbol{x}_2^m, \boldsymbol{x}_3^m, \ldots, \boldsymbol{x}_{N_g-1}^m\right]^T, \tag{Eq. 19}$$

$$Y = \left[\boldsymbol{y}_2^1, \boldsymbol{y}_3^1, \ldots, \boldsymbol{y}_{N_g-1}^1, \boldsymbol{y}_2^2, \boldsymbol{y}_3^2, \ldots, \boldsymbol{y}_{N_g-1}^2, \ldots \ldots \ldots \ldots, \boldsymbol{y}_2^{m-1}, \boldsymbol{y}_3^{m-1}, \ldots, \boldsymbol{y}_{N_g-1}^{m-1}\right]^T. \tag{Eq. 20}$$

The above data matrices establish the following system of equations

$$X = \Theta\Phi, \quad \Theta = \boldsymbol{F}(Y). \tag{Eq. 21}$$

The coefficient matrix (operator) $\Phi$ is obtained by performing a least-squares regression on both sides of the system of equations given by Eq. 21, thus,

$$\Phi = \Theta^\dagger X, \tag{Eq. 22}$$

where $\Theta^\dagger$ represents the Moore–Penrose pseudo-inverse of matrix $\Theta$. The schematic representation of the Phi Method's training phase, as was just described, is illustrated in Figure 2.

Once $\Phi$ is obtained, Eq. 16 is used as a time stepper that advances the entire state vector forward in time from a specified initial condition ($\boldsymbol{x}_i^1$, $i = 1, 2, \ldots, N_g$) and given the boundary conditions $\boldsymbol{x}_{i_b}^k$, where $i_b$ denotes the index of the nodes on the boundaries, e.g. $i_b = 1, N_g$ in a 1D domain.

It is noted that, in its current implementation, Phi Method does not incorporate a sparsity promoting regularization as is included in some other dynamics discovery algorithms like SINDy [44]. However, a sparse regression can be used in a similar manner to that in SINDy for example. In this way, the Phi Method algorithm would select as few terms as required from the library matrix Θ by zeroing out the coefficients of the unnecessary library terms. Sparse regression can further prevent overfitting the data, and the resulting "parsimonious" model will be more generalizable and interpretable. This can be particularly important for the cases where an a-priori knowledge of the functional forms of the nonlinear observables of the system does not exist. In such cases, the library of the observables must be large enough so as to ensure accounting for the involved nonlinearities in the system. As a result, it is necessary to find a *sparse* set of terms that can best represent the dynamics out of the constructed large library.

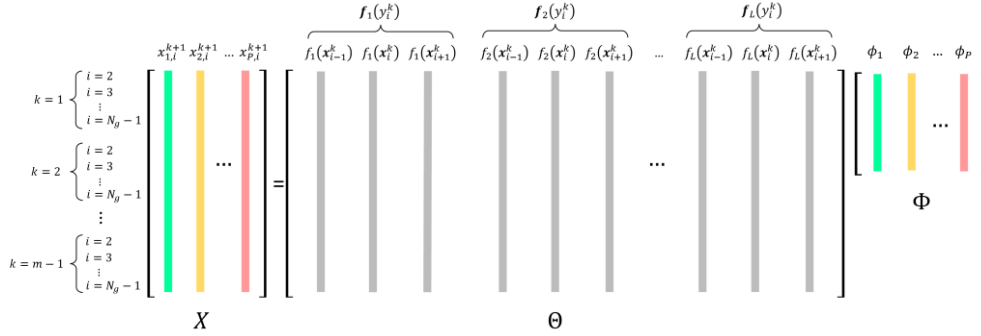

Figure 2: Schematic of the *training phase* of the Phi Method algorithm; matrix $X$ contains the state vectors at time step $k + 1$, matrix Θ represents the library of candidate terms, and matrix Φ is the coefficient matrix.

According to the descriptions provided so far, the methodology behind Phi Method sits in between the SINDy and the DMD algorithms. In this respect, similar to SINDy, Phi Method discovers a nonlinear system of differential equations by regressing on a library of terms, which are evaluated using the local values of the system's state vector data. However, Phi Method features an advancement over SINDy since, in Phi Method, we simultaneously find the system of equations and also the optimal discretization stencil. As a result, unlike SINDy which derives the continuous form of the governing differential equations [44], Phi Method provides the discretized version of the system of equations. This implies that, in Phi Method, we no longer need to pre-evaluate the spatial differential terms in the library (on a finite difference stencil, for instance) as is done in SINDy [44]. Instead, we evaluate the candidate terms on each grid node and on its neighboring nodes separately and store these values as *individual* terms in the library (Θ).

A similar approach is followed to treat the temporal differentiations. In SINDy, the time derivatives of the state variables appear on the left-hand side of the equations for which the right-hand-side terms are to be found [44]. Phi Method, however, directly relates the state variables at a subsequent time step, $t_s = k + 1$ (which represent the left-hand-side terms of the equations), to the values of the state variables at the present time step, $t_s = k$.

From the above perspective of learning the discrete time-dynamics, Phi Method shares some similarities with DMD, particularly in the case that the DMD's state vector ($X$) would represent a broader set of observables that also include nonlinear terms. Like DMD, Phi Method finds the best-fit linear operator, which maps the state vector of the present time step to the next. The key distinction between the methods lies in the fact that the Phi Method's operator (Φ) is a local operator that links the state vector at a certain spatial location to that of the immediate surrounding, i.e., the values of the state vector on the neighboring nodes. Whereas, the DMD's operator ($A$) is, by definition, global and, hence, it establishes a connection between the value of the state vector on each node to the corresponding state vector values on the entire computational domain (an entire snapshot).

In a more general sense, Phi Method shares some similarities with library-based regression techniques [54]-[57]. However, the regression in Phi Method is performed on a library of terms which is informed by numerical discretization schemes. This serves to constrain the regression, which then allows the method to be highly effective in modeling dynamics. A rather similar aim is pursued in physics-informed DMD methods [58], which are capable of significant performance improvements by imposition of physics-based symmetries or constraints. Phi Method imposes well-known discretization constraints for solving PDEs, thus, making sure that an underlying structure is respected in the regression, while enabling the exact discretization stencils to be optimally found. This significantly improves Phi Method's performance over standard library regression methods [54]-[57].

**Section 3: Phi Method verifications**

In this section, we evaluate the predictive performance and the characteristics of the Phi-Method-derived ROMs for our three adopted test cases introduced in Section 1. For each test case, we first provide an overview of the problem description and the corresponding setup that we had used to perform the ground-truth simulations. Next, we present and discuss the predictions and the behavior of the ROMs from Phi Method. For the test cases 2 and 3, i.e., the problem of fluid flow past a cylinder, and the problem of 1D azimuthal Hall thruster plasma discharge, the performance of the ROMs from Phi Method are compared against that of the OPT-DMD-derived ROMs. For these two test cases, we have additionally assessed the robustness of the Phi-Method-derived ROMs to noise in the data in the Appendix section.

### 3.1. Test case 1: Lorenz attractor

#### 3.1.1. Description of the problem setup

The Lorenz attractor is a well-known dynamical system introduced by the meteorologist Edward Lorenz in 1963 as a simplified model to study atmospheric convection [59]. The Lorenz system exhibits chaotic behavior, which is characterized by strong sensitivity to the initial conditions and the formation of a butterfly-shaped attractor. The Lorenz attractor describes the evolution of a system in three-dimensional space using a set of three coupled nonlinear ODEs (Eq. 23).

$$\frac{dx}{dt} = \sigma y - \sigma x,$$
$$\frac{dy}{dt} = \rho x - xz - y, \quad \text{(Eq. 23)}$$
$$\frac{dz}{dt} = xy - \beta z.$$

In Eq. 23, $x$, $y$, and $z$ are the state variables representing the system's evolution over time ($t$). $\sigma$, $\rho$, and $\beta$ are positive constants that characterize the behavior of the system.

Due to its simple and well-represented dynamics description, the Lorenz attractor serves as a common benchmark case for the data-driven techniques. Therefore, ahead of proceeding to higher dimensional test cases, we begin by the Lorentz attractor problem to test the Phi Method's ability in recovering the dynamics from the simulation data.

To generate the data, we have numerically integrated the Lorenz system of ODEs with the time step ($dt$) of 1 ms, and with the following values of the constant parameters: $\sigma = 10$, $\rho = 28$, and $\beta = 8/3$. The Phi Method's library of candidate terms ($\Theta$) for this test case includes the linear state variables and their quadratic functions. Hence, Eq. 16 in this example becomes

$$\begin{bmatrix} x \\ y \\ z \end{bmatrix}_{k+1}^{T} = [x \ y \ z \ x^2 \ y^2 \ z^2 \ xy \ yz \ xz]_k \Phi. \quad \text{(Eq. 24)}$$

Note that, since the state of the system for the Lorenz test case does not have a spatial distribution, i.e., there are no spatial derivatives, and, hence, the dynamics is represented by a system of ODEs, $\boldsymbol{x}$ and $\boldsymbol{y}$ in Eq. 16 are equivalent. Also, even though the dynamics is seen from Eq. 23 to not depend on the terms $x^2$, $y^2$, and $z^2$, these are included in the library to verify whether Phi Method can by itself exclude these in the learned operator.

To train the Phi Method model, the left-hand-side vector and the library matrix in Eq. 24 were evaluated using the first half of the data points collected from our numerical solution of the Lorenz problem. Following a regression on the resulting "ground-truth" training dataset, the coefficient matrix $\Phi \in \mathbb{R}^{9 \times 3}$ is determined.

#### 3.1.2. Results

Table 1 shows the values of the entries of the obtained $\Phi$ matrix. These entries correspond to the "non-zero" coefficients in Eq. 24 after the training of Phi Method. The $\Phi$ matrix entries not shown in Table 1 all had values below $10^{-5}$, which effectively amount to zero because they are about two orders of magnitude smaller than the smallest coefficient shown.

In order to compare the learned coefficients' values with the "true" coefficients, we discretized the time derivatives in Eq. 23 using first-order finite difference scheme. This leads to the discretized system of equations in Eq. 25.

$$x^k = (1 - \sigma dt) x^{k-1} + \sigma dt \, y^{k-1},$$
$$y^k = \rho dt \, x^{k-1} + (1 - dt) y^{k-1} - dt \, (xz)^{k-1}, \quad \text{(Eq. 25)}$$

$$z^k = dt(xy)^{k-1} + (1 - \beta dt)z^{k-1}.$$

Substituting the values of $dt$ and the system's parameters ($\sigma$, $\rho$, and $\beta$), the "theoretical" values of the coefficients are obtained and are presented in Table 1 as well. By comparing the Phi-Method-derived coefficients against their true values, we verify that Phi Method has effectively learned the dynamics, as evidenced by the fact that the errors in all the coefficients are below 1%.

|           | Phi Method              | True                  | Error [%] |
|-----------|-------------------------|-----------------------|-----------|
| $\Phi_{1,1}$ | $9.9018 \times 10^{-1}$ | $9.9 \times 10^{-1}$ | 0.018     |
| $\Phi_{2,1}$ | $9.9455 \times 10^{-3}$ | $1 \times 10^{-2}$    | 0.545     |
| $\Phi_{1,2}$ | $2.7955 \times 10^{-2}$ | $2.8 \times 10^{-2}$  | 0.161     |
| $\Phi_{2,2}$ | $9.9910 \times 10^{-1}$ | $9.99 \times 10^{-1}$ | 0.010     |
| $\Phi_{9,2}$ | $-9.9604 \times 10^{-4}$ | $-1 \times 10^{-3}$  | 0.396     |
| $\Phi_{3,3}$ | $9.9735 \times 10^{-1}$ | $9.973 \times 10^{-1}$ | 0.005   |
| $\Phi_{4,3}$ | $9.9318 \times 10^{-4}$ | $1 \times 10^{-3}$    | 0.682     |

Table 1: The values of the non-zero entries of the $\Phi$ matrix from Phi Method compared against the respective true values from Eq. 25. The corresponding errors are also provided. The error represents the difference between the learned and the true values of each coefficient normalized with respect to the true value.

In Figure 3, the Phi Method's prediction of the trajectory of the Lorentz attractor is compared against the true trajectory as obtained from the numerical integration of Eq. 23. This comparison further illustrates the consistency between the predicted dynamics by the Phi Method ROM and the ground-truth dynamics.

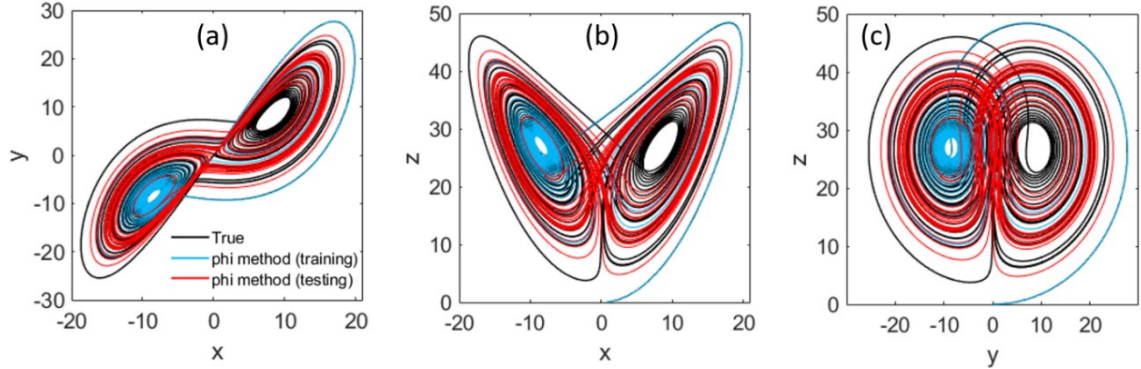

Figure 3: Comparison between the learned trajectory of a Lorenz attractor using Phi Method and the ground-truth trajectory obtained from the solution of the system of equations in Eq. 23 along (a) $y - x$, (b) $z - x$, and (c) $z - y$ planes.

### 3.2. Test case 2: 2D flow past a cylinder

#### 3.2.1. Description of the problem setup

We now extend the demonstration of Phi Method to the discovery of the spatiotemporal dependencies in a dynamical system that is governed by a PDE. Accordingly, the second test case involves a 2D unsteady fluid flow around a cylinder in the vortex-shedding regime. This example represents a classic fluid dynamics phenomenon, which is characterized by the formation of alternating vortices in the wake of a cylindrical object placed in front of a fluid stream. In this case, as the fluid flows around the cylinder, vortices are periodically shed from either side of the cylinder. This leads to the formation of a distinctive oscillatory pattern in the vorticity field (known as von Kármán vortex street), which is comprised by vortices of opposite signs. Under the assumption of an incompressible and barotropic flow, the spatiotemporal evolution of the vorticity field ($\Omega$) is given by

$$\frac{\partial \Omega}{\partial t} + (\boldsymbol{V} \cdot \boldsymbol{\nabla})\Omega = \frac{1}{Re} \nabla^2 \Omega, \tag{Eq. 26}$$

where, $Re$ represents the Reynolds number of the free stream, and $\boldsymbol{V}$ is the local flow velocity vector.

For the ground-truth data generation, we used the open-access "ViscousFlow" module [60]. ViscousFlow is a software package written in Julia language and enables simulating viscous incompressible flows around arbitrary body shapes. The 2D Cartesian domain of the simulation comprises a rectangular box in the $x - y$ plane with the dimensions $L_x = 6\ cm$ and $L_y = 4\ cm$. $x$ and $y$ represent the involved simulation directions along and perpendicular to the free stream velocity, respectively. A cylindrical object with a diameter of 1 cm is placed into the domain with its center being 1 cm from the left boundary. The left boundary serves as the flow inlet, where the Reynolds number of the freestream is 300. A time step of 1 ms is adopted, and the number of nodes along the $x$ and $y$ coordinates of the simulation domain are $N_i = 300$ and $N_j = 200$, respectively. The total simulated duration is 100 s. We output the flow properties from the simulation every 0.1 s, which resulted in a total of 1000 snapshots. The boundary conditions include a no-slip condition on the cylinder surface and the far-field conditions at the outer boundaries.

We used the collected data from the "ground-truth" simulation to train and test the Phi Method and the OPT-DMD models for the spatiotemporal evolution of the vorticity field. We discarded the initial 45 s of the simulation, which represented the transient of the system. This left us with the next 55 s duration of the simulation for the training and testing of data-driven ROMs. This 55-s duration is shown as the time interval of 0 - 55 s in the results plots of subsection 3.2.2. The first 25 s within the 55-second time window was used for the training of the models, whereas the rest was reserved to validate the forecasts obtained from the developed ROMs.

To apply OPT-DMD to the data snapshots of the vorticity field, the number of retained SVD/POD ranks in the model was 14. The OPT-DMD ROM serves as a comparison reference for the Phi-Method-derived ROM. It is noteworthy that, given the periodic behavior of the system in the present test case, DMD is expected to be a well-suited technique to provide a reduced representation of the system and to provide reliable ROMs.

For the application of Phi Method in this test case, we considered two "scenarios" in terms of the candidate library terms.

The **first scenario** involves a "linear" model, where the terms in the library are the system's state vector only. Thus, the candidate library terms include the components of the flow velocity field along the $x$ and the $y$ directions, denoted by $V_x(x,y)$ and $V_y(x,y)$, respectively, as well as the vorticity field $\Omega(x,y)$. In the case of the linear model, $\Phi_L \in R^{15 \times 1}$, and Eq. 16 becomes

$$\Omega_{i,j}\big|_{k+1} = \begin{bmatrix} \begin{bmatrix} \Omega_{i,j-1} \\ \Omega_{i-1,j} \\ \Omega_{i,j} \\ \Omega_{i+1,j} \\ \Omega_{i,j+1} \end{bmatrix}^T & \begin{bmatrix} V_{x_{i,j-1}} \\ V_{x_{i-1,j}} \\ V_{x_{i,j}} \\ V_{x_{i+1,j}} \\ V_{x_{i,j+1}} \end{bmatrix}^T & \begin{bmatrix} V_{y_{i,j-1}} \\ V_{y_{i-1,j}} \\ V_{y_{i,j}} \\ V_{y_{i+1,j}} \\ V_{y_{i,j+1}} \end{bmatrix}^T \end{bmatrix}_k \Phi_L. \tag{Eq. 27}$$

In Eq. 27, each of the inner vectors of the matrix on the right-hand side characterizes a state variable on the node $(i,j)$ and on its neighboring nodes. These vectors were represented by $y_i$ in Eq. 16. Note that, for a linear model, $f_l(y_i) = y_i$, and $l = 1, 2, \ldots, P$, where $P$ is the total number of state variables.

The **second scenario** involves a "nonlinear" model, which, in addition to the linear terms – $V_x(x,y)$, $V_y(x,y)$, and $\Omega(x,y)$ – includes some quadratic nonlinear functions of the state vector as well, namely, $\Omega V_x(x,y)$ and $\Omega V_y(x,y)$. Choosing these specific nonlinear terms have been in accordance with the terms that appear in the dynamics PDE in Eq. 26. The impact of considering a broader set of terms in the library that includes all the quadratic functions of the state variables will be discussed in the subsection 3.2.2. Eq. 16 for the nonlinear model takes the following form

$$\Omega_{i,j}\big|_{k+1} = \begin{bmatrix} \begin{bmatrix} \Omega_{i,j-1} \\ \Omega_{i-1,j} \\ \Omega_{i,j} \\ \Omega_{i+1,j} \\ \Omega_{i,j+1} \end{bmatrix}^T & \begin{bmatrix} V_{x_{i,j-1}} \\ V_{x_{i-1,j}} \\ V_{x_{i,j}} \\ V_{x_{i+1,j}} \\ V_{x_{i,j+1}} \end{bmatrix}^T & \begin{bmatrix} V_{y_{i,j-1}} \\ V_{y_{i-1,j}} \\ V_{y_{i,j}} \\ V_{y_{i+1,j}} \\ V_{y_{i,j+1}} \end{bmatrix}^T & \begin{bmatrix} \Omega V_{x_{i,j-1}} \\ \Omega V_{x_{i-1,j}} \\ \Omega V_{x_{i,j}} \\ \Omega V_{x_{i+1,j}} \\ \Omega V_{x_{i,j+1}} \end{bmatrix}^T & \begin{bmatrix} \Omega V_{y_{i,j-1}} \\ \Omega V_{y_{i-1,j}} \\ \Omega V_{y_{i,j}} \\ \Omega V_{y_{i+1,j}} \\ \Omega V_{y_{i,j+1}} \end{bmatrix}^T \end{bmatrix}_k \Phi_{NL}, \tag{Eq. 28}$$

where $\Phi_{NL} \in R^{25 \times 1}$.

In order to avoid possible numerical issues due to the discrepancy in the order of magnitudes of the different state variables (flow properties), each state variable was normalized with respect to its maximum value over the entire spatiotemporal data. The Φ operators (matrices) for the "linear" and the "nonlinear" models, i.e., $\Phi_L$ and $\Phi_{NL}$, respectively, were obtained by performing regression on the data matrices in Eqs. 27 and 28. The data matrices were constructed by evaluating the left-hand sides of the Eq. 27 and the Eq. 28 along with their respective library terms over a sufficiently large number of data points during the training interval.

### 3.2.2. Results

The obtained linear and nonlinear ROMs from Phi Method were used to reconstruct the vorticity field data within the training window, starting from a given initial condition, and to forecast the vorticity field beyond the training dataset. The results in terms of the time evolution of the spatially averaged vorticity ($\Omega_{mean}$) and the local value of the vorticity field ($\Omega_{mid}$) at the center of the domain ($x = 3$ cm and $y = 2$ cm) are presented in Figure 4. The corresponding predicted traces from the OPT-DMD ROM are superimposed on the plots for comparison.

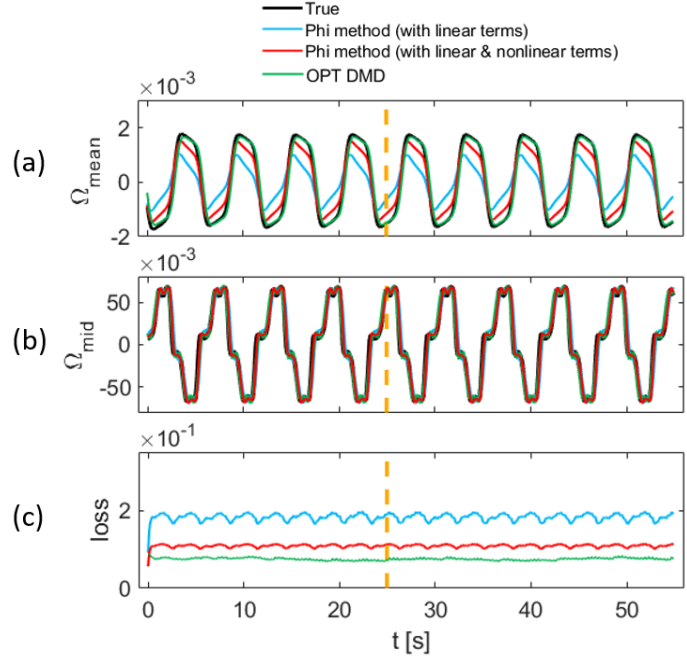

Figure 4: Comparison of the linear and nonlinear ROMs from Phi Method and the ROM from OPT-DMD against the ground-truth simulation for the test case 2; time evolution of (a) spatially averaged normalized vorticity ($\Omega_{mean}$), (b) local value of the normalized vorticity at the mid-location within the simulation domain ($\Omega_{mid}$), and (c) the loss factor over the entire domain. Dashed yellow lines indicate the end of the training interval.

The instantaneous loss of the predictions from the Phi Method ROMs and the OPT-DMD ROM with respect to the original ground-truth data are calculated according to Eq. 29. The time evolution of the loss factor corresponding to each DD model is also plotted in Figure 4.

$$Loss = \frac{\|S_k^{true} - S_k^{ROM}\|_F}{\|S_k^{true}\|_F}, \quad \|S\|_F \equiv \sqrt{\sum_{i=1}^{N_i}\sum_{j=1}^{N_j}|s_{i,j}|^2}.$$
(Eq. 29)

In Eq. 29, $\|S_k\|_F$ represents the Frobenius norm of a data snapshot $S$ at the time step $k$. $S_k^{true}$ is the original (ground-truth) snapshot from the simulation, and $S_k^{ROM}$ is the predicted snapshot from the derived ROM. Also, $N_i$ and $N_j$ are the number of grid nodes along the $x$ and the $y$ directions of the simulation, $m_t$ is the number of snapshots during the test interval, and $s_{i,j}$ represents the value of the state variable of interest on the node $(i,j)$.

From Figure 4, we see that even the Phi Method's linear model provides a remarkably good approximation of the original data, which is particularly evident when looking at the time evolution plot of the local value of the vorticity. This point is further evidenced by comparing the predicted 2D snapshots of the vorticity field from the Phi Method linear ROM against the true snapshots as are shown in Figure 5.

In terms of the spatially averaged vorticity ($\Omega_{mean}$), however, Figure 4(a) shows that the inclusion of the nonlinear terms leads to improvement in the accuracy of the Phi Method model. The time evolution of the loss factor in Figure 4(c) shows quantitatively the degree of improvement achieved by the Phi Method nonlinear ROM. In particular, the loss factor for the nonlinear model is almost half of that for the linear model.

Another interesting observation is that the Phi Method linear model exhibits the largest loss when the absolute value of the average vorticity reaches the maximum in each cycle. Contrarily, the linear model's loss is minimum when the average vorticity is close to zero. A similar variation in the loss factor is also noticed for the nonlinear ROM, albeit with a smaller amplitude. The loss associated with the OPT-DMD model, nevertheless, remains nearly constant throughout the entire duration of the signal. The mean loss of the Phi Method nonlinear model is about 1.4 times that of the OPT-DMD model. Finally, for all the models, the level of loss is seen to remain consistent between the training and the test intervals.

In Figure 5, we have shown the predicted 2D snapshots of the normalized vorticity field from each derived ROM at three sample time instants beyond the training duration. The true 2D snapshots from the simulation are also shown for reference. The overall patterns in the snapshots from all the models are almost indistinguishable from their true counterparts. To illustrate, however, the subtle discrepancies not readily apparent in the snapshots of Figure 5, Figure 6 presents the absolute-difference plots between the true snapshots and the corresponding predicted ones ($S_k^{true} - S_k^{ROM}$) at the three sample time instants. From this figure, it is evident that the error of the Phi Method linear ROM is the highest among the models presented, whereas the errors associated with the Phi Method nonlinear model and the OPT-DMD ROM are within a comparable range.

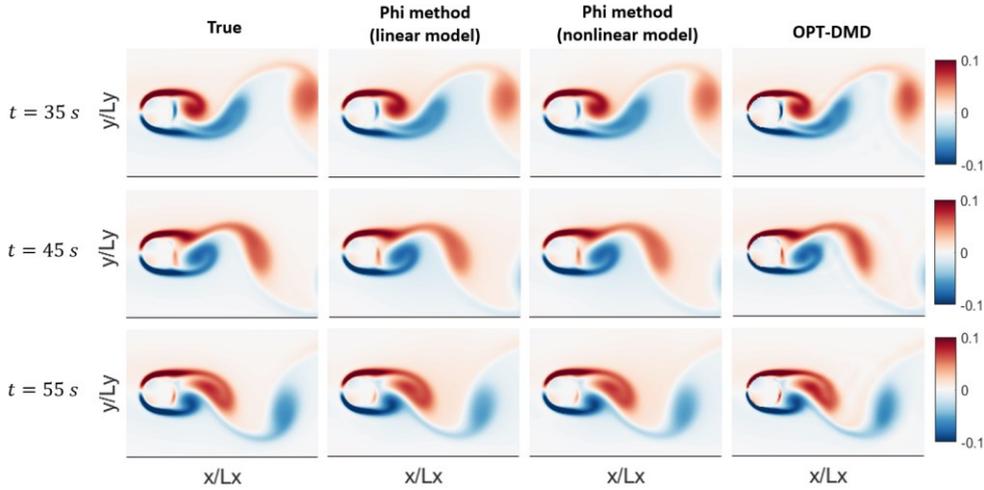

Figure 5: Comparison of the linear and the nonlinear ROMs from Phi Method and the ROM from OPT-DMD against the ground-truth simulation for the test case 2; snapshots of the normalized vorticity field at three different time instances within the test interval.

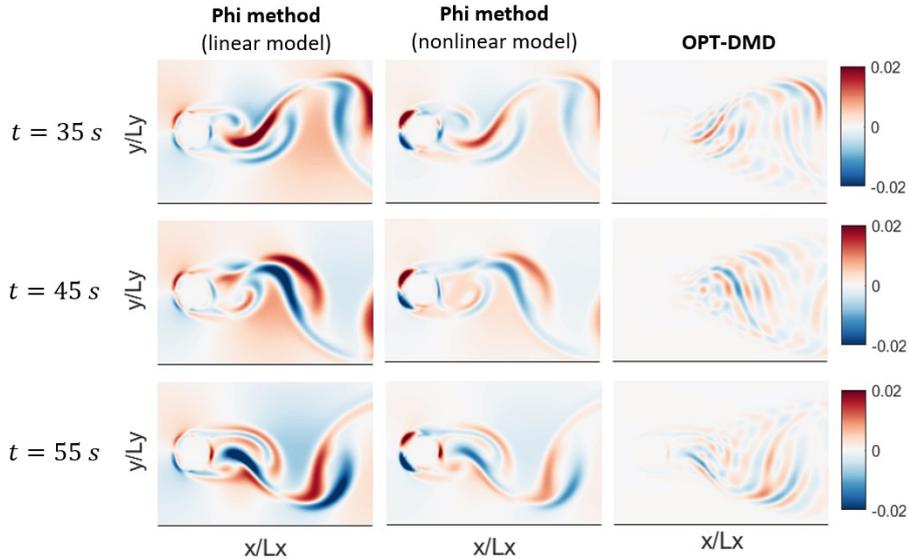

Figure 6: Comparison of the linear and the nonlinear ROMs from Phi Method and the ROM from OPT-DMD for the test case 2; plots of the absolute difference between the true and the predicted normalized vorticity snapshots at three different time instants during the test interval.

Also, we notice from Figure 6 that the patterns in the difference plots corresponding to the Phi Method linear and nonlinear ROMs are different from the difference-plot patterns for the OPT-DMD model. For OPT-DMD, the non-zero error is confined to the downstream wake region, whereas for Phi Method, the error extends across the entire flow field. This distinction arises from the inherent differences between the Phi Method and the OPT-DMD models. OPT-DMD yields a "global" model, and its error represents the sum of the discarded DMD modes (i.e., beyond rank 14, where we truncated the SVD/POD expansion). In this regard, although the upstream and the lateral regions of the flow are fully represented by the leading DMD modes that we retained, the wake exhibits higher dimensionality which requires a larger number of modes for a complete representation. The Phi Method models, on the contrary, are local operators that best fit the dynamics across the entire flow field because a single PDE is expected to govern the dynamics everywhere in the flow. Hence, for the case of Phi Method, the error arises from a dynamical term not having been entirely captured by the model, thereby resulting in prediction errors throughout the entire domain.

Moving on, we have so far used the data from the quasi-steady oscillatory state of the system for the training of the data-driven models, excluding the initial transient behavior from the dataset. However, there might be scenarios in which the dynamics does not feature a steady state, or the system takes a long time to reach a quasi-steady state. In these scenarios, our only option would be to train a data-driven model using the transient data of the system. Accordingly, to evaluate Phi Method's capability to learn the dynamics from transient data, we trained both the linear and the nonlinear models on the initial 0-25 s duration of the simulation time during which the flow exhibits a highly transient behavior. This 25-second time window falls within the initial 45-second interval that we had discarded before for the results discussed through Figure 4 to Figure 6.

The outcomes of the above exercise are provided in Figure 7 in terms of the time evolutions of the predicted vorticity field signals from both the linear and the nonlinear Phi Method models as well as the time evolution of the associated loss factors. It is observed that the models' predictions of the mean and the local vorticity values are reasonably consistent with the original (true) data. Furthermore, the mean loss of the models when trained on the transient-state data is only about 1.7 times that of the corresponding models when trained on the steady-state data. This verifies that Phi Method is indeed able to identify the optimal Φ operator from the transient data as well. It is highlighted that, since the characterization of the transient phenomena is a well-known limitation of DMD-based (POD-based) approaches [32], OPT-DMD could only be reliably trained on the quasi-steady data and, hence, the predictions of the OPT-DMD model are not shown in the plots of Figure 7.

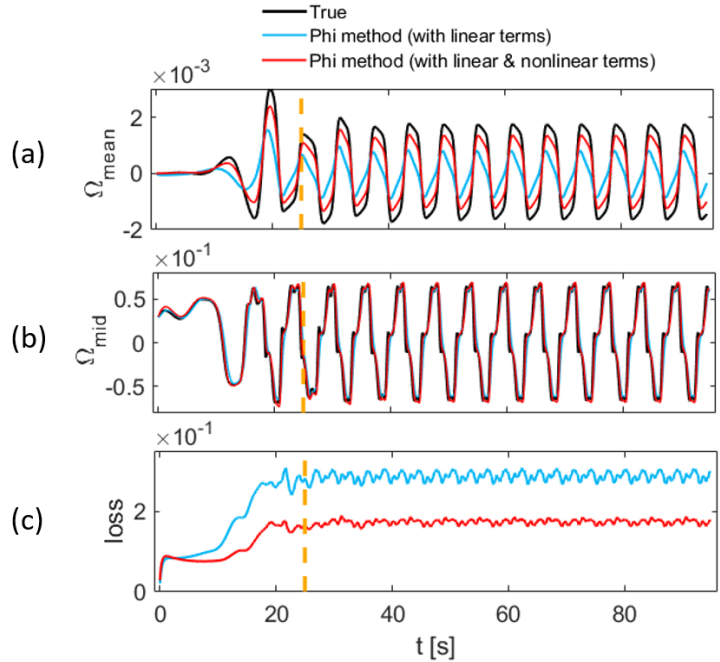

Figure 7: Comparison of the linear and nonlinear Phi Method models trained on the transient-state data against the ground-truth simulation for the test case 2; time evolutions of (a) spatially averaged normalized vorticity, (b) local value of the normalized vorticity at the mid-location within the simulation domain, and (c) loss factor over the entire domain. Dashed yellow lines indicate the end of the training interval.

As a final discussion in this section, we have provided in Figure 8 the visualizations of the $\Phi_L$ and the $\Phi_{NL}$ matrices (operators) that correspond, respectively, to the linear and the nonlinear Phi Method ROMs. We have visualized the $\Phi_L$ and the $\Phi_{NL}$ matrices from the Phi Method training on both the steady-state (45 – 65 s) and the transient-state (0 – 25 s) data. It is noted that, although the $\Phi$s are column matrices, we have illustrated a rectangular rearrangement of the matrices in Figure 8.

Apart from the $\Phi_L$ and the $\Phi_{NL}$ matrices that were discussed up to this point, a $\Phi$ matrix corresponding to another nonlinear model is also visualized. This new nonlinear model incorporates extra terms, namely, $V_x^2(x,y)$ and $V_y^2(x,y)$, in addition to the terms that were present in the so-far-analyzed nonlinear model (Eq. 28). The $\Phi$ matrix associated with this new model ($\Phi_{NL,ET}$) is also visualized from the model's training on the steady-state and the transient-state data.

The matrices $\Phi_L$, $\Phi_{NL}$, and $\Phi_{NL,ET}$ correspond to the Phi Method ROMs with increasing levels of complexity as is determined by the number of candidate terms in the respective libraries of the models.

We would highlight several points from the comparisons among the various $\Phi$ matrices in Figure 8: first, at each complexity level (each row), the main characteristics of the $\Phi$ operators are shared between the models trained on the steady or the transient data. This includes the structure of the operator, which represents the dominant terms identified and the discretization stencil.

Second, across the complexity levels (column-wise comparison), the coefficients corresponding to each specific term are very similar. This provides confidence that the addition of extra terms in the library, which may not be necessary to describe the dynamics, will not significantly impact the coefficients found for the essential and dominant terms. In practice, to achieve a balance between model's fidelity and complexity, we can plot the Pareto front of the loss factor vs the number of involved dynamical terms and identify the optimum model.

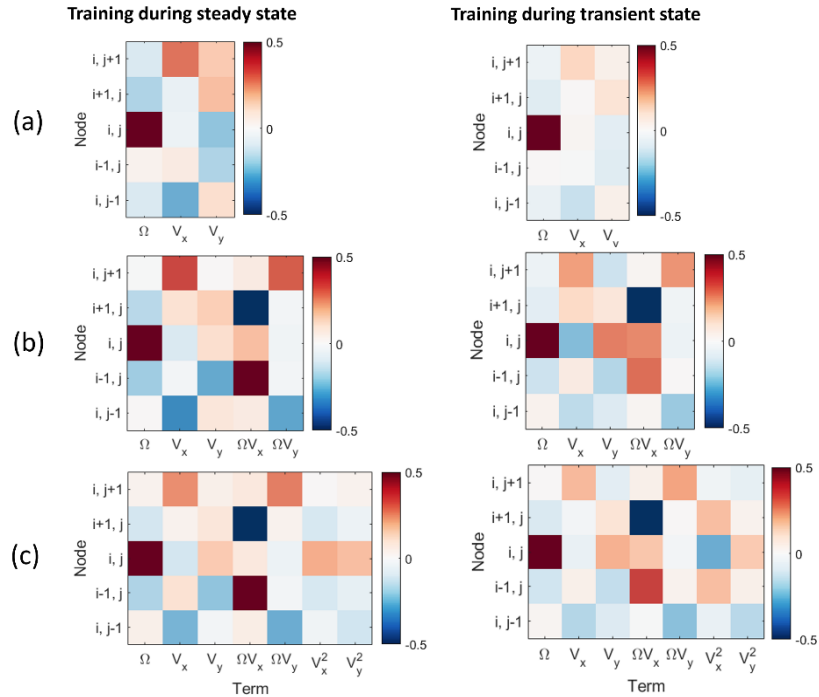

Figure 8: Rearranged normalized representations of various coefficient matrices (operators, $\Phi$) from the Phi method's application to the fluid flow test case; (a) visualization of the $\Phi_L$ matrix, (b) visualization of the $\Phi_{NL}$ matrix, and (c) visualization of the $\Phi_{NL,ET}$ matrix that represents a larger library of terms compared to the ROM corresponding to $\Phi_{NL}$. (**left column**) models' training performed on the steady-state data, (**right column**) models' training performed on the transient-state data. $i$ and $j$ represent the node's indices along the $x$ and the $y$ directions, respectively.

Third, it is emphasized that, by looking at a $\Phi$ matrix, we can infer information about the discretized dynamics in terms of the relative importance of the candidate terms and their optimal discretization stencil. To be able to compare the obtained data-driven coefficients with their analytical values, we rewrite Eq. 26 in the following form

$$\frac{\partial \Omega}{\partial t} = -\frac{\partial \Omega V_x}{\partial x} - \frac{\partial \Omega V_y}{\partial y} + \Omega \frac{\partial V_x}{\partial x} + \Omega \frac{\partial V_y}{\partial y} + \frac{1}{Re}\left(\frac{\partial^2 \Omega}{\partial x^2} + \frac{\partial^2 \Omega}{\partial y^2}\right). \quad \text{(Eq. 30)}$$

According to Eq. 30, the terms that appear in the dynamics include $\Omega$, $\Omega V_x$, and $\Omega V_y$, which are also the terms with the dominant coefficients as can be seen from Figure 8.

Discretizing Eq. 30 using first-order Euler scheme in time and second-order finite difference in space, we obtain

$$\Omega_{i,j}^{k+1} = \left(1 - \frac{2\Delta t}{Re}\left(\frac{1}{\Delta x^2} + \frac{1}{\Delta y^2}\right)\right)\Omega_{i,j}^k + \frac{\Delta t}{Re}\left(\frac{1}{\Delta x^2}(\Omega_{i+1,j}^k + \Omega_{i-1,j}^k) + \frac{1}{\Delta y^2}(\Omega_{i,j+1}^k + \Omega_{i,j-1}^k)\right)$$
$$- \frac{\Delta t}{2\Delta x}\left((\Omega V_x)_{i+1,j}^k - (\Omega V_x)_{i-1,j}^k\right) - \frac{\Delta t}{2\Delta y}\left((\Omega V_y)_{i,j+1}^k - (\Omega V_y)_{i,j-1}^k\right) \quad \text{(Eq. 31)}$$
$$+ \frac{\Delta t}{2\Delta x}\Omega_{i,j}^k \left(V_{x_{i+1,j}}^k - V_{x_{i-1,j}}^k\right) + \frac{\Delta t}{2\Delta y}\Omega_{i,j}^k \left(V_{y_{i,j+1}}^k - V_{y_{i,j-1}}^k\right)$$

Referring to the relation in Eq. 31, we point out that, in the current implementation of Phi Method, all constituent variables of a library term are evaluated on the same node. Accordingly, the terms such as $\Omega \frac{\partial V_x}{\partial x}$ and $\Omega \frac{\partial V_y}{\partial y}$, which appear as "cross-node" terms (i.e., for example, $\Omega_{i,j}^k V_{x_{i+1,j}}^k$, $\Omega_{i,j}^k V_{x_{i-1,j}}^k$, $\Omega_{i,j}^k V_{y_{i,j+1}}^k$, and $\Omega_{i,j}^k V_{y_{i,j-1}}^k$) in the discretized relation of Eq. 31 are not represented in a Phi Method model. As a result of this, a direct comparison between the learned coefficients and their analytical counterparts is not straightforward. In fact, for the fluid flow test case, Phi Method is effectively finding an *alternative* model that can optimally represent the data. Despite this point, a symmetrical pattern (with respect to node $(i,j)$) can be overall observed in the $\Phi$ matrices depicted in Figure 8. This symmetrical pattern is reminiscent of the finite difference stencil, which has indeed been the underlying discretization stencil that was used for the simulation that generated the training and the test data.

### 3.3. Test case 3: 1D azimuthal plasma configuration

#### 3.3.1. Description of the problem setup

The setup of the 1D azimuthal plasma simulation is adopted from Ref. [61]. The test case corresponds to the problem of fluctuations in plasma properties along the azimuthal coordinate of an E × B plasma discharge. An E × B discharge refers to a type of plasma configuration where the electric field ($E$) and the magnetic field ($B$) are mutually perpendicular. In this plasma geometry, the azimuthal direction ($z$) is perpendicular to both the electric and the magnetic field, hence, along the E × B direction. In our test case, the external electric field is along the axial ($x$) direction, and the magnetic field is along the radial ($y$) direction. The physics of the problem involves the development, growth, and saturation of the Electron Cyclotron Drift Instability (ECDI) waves, which are of relatively high-frequency (from 1 – 10 MHz) and short wavelength (on the order of 1 mm) [3][61] and exhibit nonlinear behaviors during the growth and saturation phases [3][61].

For the data generation, we ran a 1D electrostatic PIC simulation using IPPL-1D code [25][26]. The simulation solves the Poisson's equation along the $z$-direction using a direct tridiagonal matrix solve algorithm (Thomas algorithm) given the charge distribution from the plasma species (electrons and ions), and in turn computes the self-consistent azimuthal electric field. The positions and the velocities of the plasma particles are updated at each time step of the simulation according to the electromagnetic forces at the particles' location [18][19]. The details of the simulation setup and conditions are described in the following.

The simulated domain represents a 0.5 cm-long extent of the azimuthal coordinate of a Hall thruster, which is an industrially important E × B plasma technology. A uniform radial magnetic field ($B_y$) with an intensity of 20 mT is applied. Also, an axial electric field ($E_x$) with the magnitude of $2 \times 10^4\ Vm^{-1}$ is imposed with a uniform distribution across the domain. The simulation's initialization involves sampling the electrons and the ions from their initial Maxwellian distributions at the temperatures of 2 eV and 0.1 eV, respectively. The sampled particles are then loaded throughout the domain with a uniform density of $1 \times 10^{17}\ m^{-3}$. The domain is discretized along the azimuthal ($z$) coordinate using $N_g = N_i = 100$ nodes. At the simulation's start, the number of macroparticles per cell for either the electron or the ion species is 200. The time step is $5 \times 10^{-12} s$, and the total simulated time is 10 $\mu s$. The plasma properties are averaged over every 500-time steps (2.5 $ns$) and recorded in our dataset. Thus, the total number of plasma snapshots is 4000.

A fictitious axial extent with the length of 1 cm is considered. The axial extent is included to provide a means to represent the axial convection of the instability waves, which is important to properly capture their saturation [61]. This axial extent also enables limiting the growth of the charged particles' energy [61]. The electrons or the ions that leave the domain from one of the axial boundaries are resampled from their initial distributions and are reloaded at the opposite axial end of the domain with a random azimuthal position.

Along the azimuthal direction, a periodic boundary condition is applied on the particles. No boundary condition is applied on the particles along the radial coordinate, and particles move freely in this direction. Note that, despite assuming a finite axial length, the simulation still represents a 1D problem, since the Poisson's equation is only solved along the azimuthal coordinate. An azimuthally periodic solution for the plasma potential is ensured by imposing zero-volt Dirichlet conditions at both azimuthal ends of the domain.

The test case is collisionless. This assumption is to simplify the involved dynamics and, hence, to limit the number of candidate terms that may need to be accounted for in the Phi Method's library.

For the present test case, we aim to use Phi Method to discover a system of discretized PDEs that describes the spatiotemporal evolution of multiple plasma properties simultaneously. These plasma properties, which constitute the output vector of the Phi Method ROM, are the electron number density ($n_e$), the electrons' azimuthal and axial drift velocity components ($V_{d,ez}$ and $V_{d,ex}$), and the azimuthal electric field ($E_z$).

The library of candidate terms consists of the state variables in the output vector themselves (the above quantities) in addition to a few other nonlinear functions of them, namely, $n_e V_{d,ez}$, $n_e V_{d,ez}^2$, $n_e E_z$, and $n_e T_{ez}$. The last term, $n_e T_{ez}$, represents the electron pressure, $\Pi_{zz} = K n_e T_{ez}$, where $K$ is the Boltzmann constant, and $T_{ez}$ is the azimuthal electron temperature.

Accordingly, for this test case 3, $\Phi \in R^{24 \times 4}$, and Eq. 16 is expressed as

$$\begin{bmatrix} n_{e_i} \\ V_{d,ez_i} \\ E_{z_i} \\ V_{d,ex_i} \end{bmatrix}_{k+1}^T = \begin{bmatrix} \begin{bmatrix} n_{e_{i-1}} \\ n_{e_i} \\ n_{e_{i+1}} \end{bmatrix}^T \begin{bmatrix} V_{d,ez_{i-1}} \\ V_{d,ez_i} \\ V_{d,ez_{i+1}} \end{bmatrix}^T \begin{bmatrix} E_{z_{i-1}} \\ E_{z_i} \\ E_{z_{i+1}} \end{bmatrix}^T \begin{bmatrix} V_{d,ex_{i-1}} \\ V_{d,ex_i} \\ V_{d,ex_{i+1}} \end{bmatrix}^T \begin{bmatrix} (n_e V_{d,ez})_{i-1} \\ (n_e V_{d,ez})_i \\ (n_e V_{d,ez})_{i+1} \end{bmatrix}^T \begin{bmatrix} (n_e V_{d,ez}^2)_{i-1} \\ (n_e V_{d,ez}^2)_i \\ (n_e V_{d,ez}^2)_{i+1} \end{bmatrix}^T \begin{bmatrix} (n_e E_z)_{i-1} \\ (n_e E_z)_i \\ (n_e E_z)_{i+1} \end{bmatrix}^T \begin{bmatrix} (n_e T_{ez})_{i-1} \\ (n_e T_{ez})_i \\ (n_e T_{ez})_{i+1} \end{bmatrix}^T \end{bmatrix}_k \Phi. \quad \text{(Eq. 32)}$$

The choice of the considered dynamical library terms in Eq. 32 is informed by the conservation equations for the electrons, namely, the continuity equation (Eq. 33) and the 1D momentum equations along the axial and the azimuthal directions (Eq. 34 and Eq. 35)

$$\frac{\partial n_e}{\partial t} = -\frac{\partial n_e V_{d,ez}}{\partial z}, \quad \text{(Eq. 33)}$$

$$q B_y n_e V_{d,ez} = m_e \frac{\partial (n_e V_{d,ex})}{\partial t} + m_e \frac{\partial (n_e V_{d,ex} V_{d,ez})}{\partial z} - q n_e E_x + \frac{\partial \Pi_{xz}}{\partial z}, \quad \text{(Eq. 34)}$$

$$-q B_y n_e V_{d,ex} = m_e \frac{\partial (n_e V_{d,ez})}{\partial t} + m_e \frac{\partial (n_e V_{d,ez}^2)}{\partial z} - q n_e E_z + \frac{\partial \Pi_{zz}}{\partial z}. \quad \text{(Eq. 35)}$$

In the above equations, $q$ and $m_e$ are the elementary charge and the electron's mass, respectively. $\Pi_{zz}$ and $\Pi_{xz}$ are the pressure tensor components ($\Pi_{ij}$), which are defined as the second moments of the electrons' velocity distribution function ($f_e(v)$) according to

$$\Pi_{ij} = m_e \int_{-\infty}^{\infty} (v - V_{d,ei}(v - V_{d,ej}) f_e(v) d^3 v. \quad \text{(Eq. 36)}$$

Similar to what was done for the test case 2, the data corresponding to each plasma variable is normalized here with respect to its maximum value. This ensures the order-of-magnitude consistency across the involved terms. The initial 2 $\mu s$ of the data is disregarded as this time period represents the transient phase of the system corresponding to the excitation and the growth of the ECDI. This was to enable the application of OPT-DMD to the data for the purpose of comparisons between the OPT-DMD-derived and the Phi-Method-derived ROMs.

To train the Phi Method model, the output vector and the terms in the library matrix were evaluated on several data points during the training interval that spanned from 2 to 6 $\mu s$. Like in the previous test cases, the $\Phi$ matrix was obtained following a regression on the resulting data matrices. The rest of the simulation data corresponding to the timeframe of 6 – 10 $\mu s$ were reserved for testing the model.

With regard to the OPT-DMD model, in order for it to be comparable with the Phi Method ROM, we used the same terms in the Phi Method's library to construct the augmented state vectors ($\mathcal{X}_k$) for OPT-DMD (Section 2.1). The $\mathcal{X}_k$s comprise the rearranged snapshots of the normalized state variables, in addition to their relevant nonlinear observables, according to Eq. 13. OPT-DMD was applied to the augmented matrix $X$ (Eq. 1), which contains the state vectors $\mathcal{X}_k$, with $k \in [800, 2400)$ corresponding to the training interval of $2-6\ \mu s$. The leading 25 spatiotemporal DMD modes were retained for the OPT-DMD ROM.

*3.3.2. Results*

The predictive performance of the ROMs derived from Phi Method and OPT-DMD are analyzed in this subsection.

Figure 9 shows the time evolution signals of the output state variables ($n_e$, $V_{de,z}$, $E_z$, and $V_{de,x}$) at the mid-location of the domain from the Phi Method and the OPT-DMD ROMs. The time variations of these variables are reconstructed during the training interval and are forecasted within the test period. The predicted signals are compared against the ground-truth from the PIC simulation. Additionally, in Figure 10, the predictions of the data-driven ROMs regarding the full spatiotemporal evolution of the output plasma state variables are provided and compared against the original data.

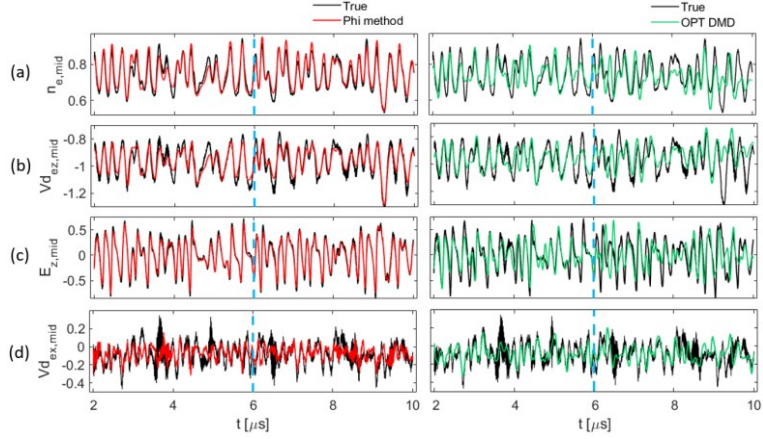

Figure 9: Comparison of the predictions from the Phi Method (left column) and the OPT-DMD (right column) ROMs against the ground-truth data for the test case 3; time evolutions of the local values at the mid-location within the simulation domain of (a) normalized electron number density ($n_e$), (b) normalized electrons' azimuthal drift velocity ($V_{d,ez}$), (c) normalized azimuthal electric field ($E_z$), and (d) normalized electrons' axial drift velocity ($V_{d,ex}$). Dashed blue lines indicate the end of the training interval.

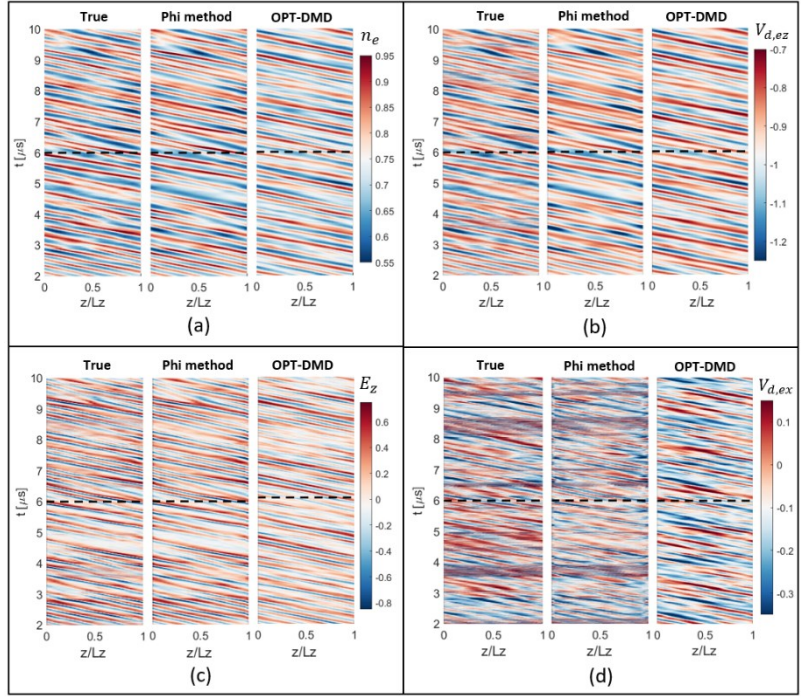

Figure 10: Comparison of the predictions from the Phi Method and the OPT-DMD ROMs against the ground-truth data for the test case 3; spatiotemporal evolution plots of (a) normalized electron number density ($n_e$), (b) normalized electrons' azimuthal drift velocity ($V_{d,ez}$), (c) normalized azimuthal electric field ($E_z$), and (d) normalized electrons' axial drift velocity ($V_{d,ex}$). Dashed black lines indicate the end of the training interval.

We would highlight that, in advancing the Phi Method ROM from one time step to the next, all the *input* variables named in subsection 3.3.1, i.e., the library terms in Eq. 32, were obtained from the model itself except for the $n_e T_{ez}$, which was fed into the model from the PIC simulation data. With this exception, the model-predicted values were used to update all the other library terms for the prediction of the *output* state vector at the subsequent time step.

Besides, to reflect the azimuthal periodicity in the data, we used Eq. 16 with the following redefinition of $\boldsymbol{y}$ vectors at the boundary nodes

$$\boldsymbol{y}_i^k = \begin{bmatrix} \boldsymbol{x}_{N_i-1}^k \\ \boldsymbol{x}_i^k \\ \boldsymbol{x}_{i+1}^k \end{bmatrix}^T, \qquad i = 1, \tag{Eq. 37}$$

$$\boldsymbol{y}_i^k = \begin{bmatrix} \boldsymbol{x}_{i-1}^k \\ \boldsymbol{x}_i^k \\ \boldsymbol{x}_2^k \end{bmatrix}^T, \qquad i = N_i. \tag{Eq. 38}$$

Therefore, the boundary values are computed within the model itself given the initial conditions.

The OPT-DMD ROM is fully self-consistent, i.e., the values of the complete state vector ($X$) are computed within the model at each time step.

It is important to note that the ability of a data-driven ROM to provide reliable predictions of the *local* values of the state variables of a plasma system is a challenging task. This is because the local values of the plasma properties are affected by an array of intricate plasma phenomena and interactions. In this respect, Figure 9 and Figure 10 clearly illustrate that Phi Method reliably manages to handle this task as it is evidenced by the remarkable agreement between the Phi Method ROM's predictions with the corresponding ground-truth data.

Furthermore, when evaluating the performance of the OPT-DMD model, one must consider its inherent self-consistency as well as the challenging aspect of predicting the system's dynamics in the absence of a fully periodic behavior. Despite these factors, OPT-DMD is seen in Figure 9 and Figure 10 to provide a reasonably accurate prediction of the system's state vector.

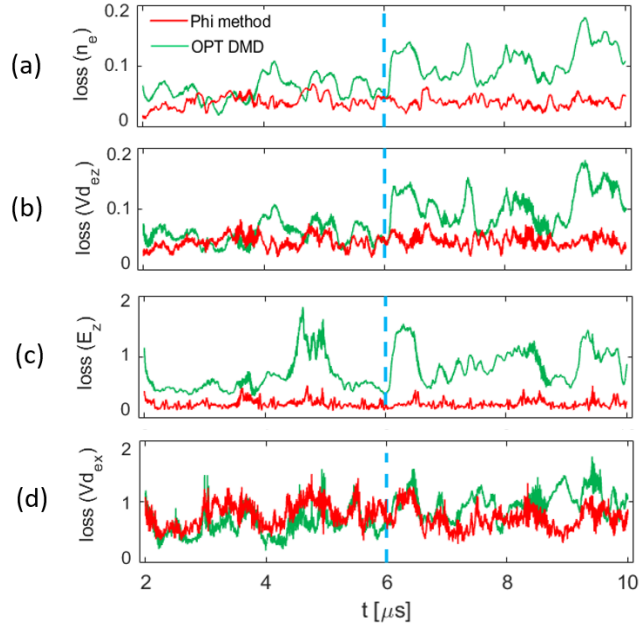

Figure 11: Comparison of the ROMs from Phi Method and OPT-DMD for the test case 3; time evolutions of the loss factor calculated over the entire domain corresponding to (a) normalized electron number density ($n_e$), (b) normalized electrons' azimuthal drift velocity ($V_{d,ez}$), (c) normalized azimuthal electric field ($E_z$), and (d) normalized electrons' axial drift velocity ($V_{d,ex}$). Dashed blue lines indicate the end of the training interval.

For a more quantitative comparison between the accuracy of predictions from the Phi Method and OPT-DMD ROMs, we refer to Figure 11. This figure shows the time variations of the loss factor associated with the data-driven models' predictions of the various plasma properties. The loss factor is computed with respect to ground-truth data and over the entire domain in accordance with Eq. 29.

From the plots in Figure 11, we observe that, first of all, the mean of the loss over time corresponding to the Phi Method ROM is smaller than the mean loss corresponding to the OPT-DMD ROM for all the plasma properties except for the $V_{d,ex}$ (Figure 11(d)), where the models from both methods exhibit comparable mean loss values.

Second, the Phi Method's loss in the forecasting window remains at a similar level to that during the training. In contrast, the OPT-DMD's predictions experience a slight increase in loss factor during the test window. This difference in the loss factor behavior between the two models stems from their inherently distinct nature as was also explained in subsection 3.2.2. The fundamental assumption of Phi Method is that the dynamics of the system can be characterized by a unique set of (partial) differential equations throughout the evolution of the system. Once these equations are identified through the best-fit operator $\Phi$, they are presumed to effectively capture the system's evolution, whether reconstructing the observed data during the training phase or forecasting the unseen data beyond the training interval. OPT-DMD also assumes that the spatial modes and their dynamics remain constant throughout time. However, while the assumption of the constancy of PDEs is a realistic assumption for physical systems, the time-invariability of the DMD modes is only strictly valid for a fully periodic system. Accordingly, due to slight variations in the existing DMD's spatiotemporal modes from the training to the test interval, the observed increase in the OPT-DMD's loss beyond the training point is expected.

Third, we observe from Figure 11 that some plasma properties are predicted with a higher accuracy than the others. Furthermore, the relative accuracy between the Phi Method and the OPT-DMD ROMs varies across different plasma properties. In particular, the loss associated with the Phi Method's predictions of the $n_e$ is the smallest among the various state properties and remains below 0.05 at all times. The Phi Method and the OPT-DMD models show the largest disparity in their loss levels toward predicting the $E_z$, with the OPT-DMD's loss being more than three times that of the Phi Method's. In addition, both models exhibit nearly identical loss levels (below 1 on average) for the $V_{d,ex}$. The loss factor for the $V_{d,ex}$ represents the highest value relative to the loss factors associated with the other predicted plasma properties.

*3.3.2.1. Impacts of the size of the library on the Phi Method ROM*

In this subsection, we study the impacts that the size of the library of candidate terms may have on Phi Method, assessing the extent to which the number of candidate library terms can affect the learned operator $\Phi$.

For the purpose of the above analysis, we focused on a particular plasma property, namely, the electrons' axial current density ($J_{ex} = -qn_e V_{d,ex}$). This quantity is of particular interest since it represents the aggregate contributions of various effects to the phenomenon of electrons' transport across the magnetic fields. The lack of an accurate prediction of this quantity has constrained the self-consistency of the fluid-based plasma simulations so far.

We have trained two distinct Phi Method models for the $J_{ex}$ property using the simulation data, with each model being represented by a different set of library terms. The first model is characterized by a small library of terms that includes only the four essential terms that appear in the governing PDE (Eq. 35). These are $n_e V_{d,ez}$, $n_e E_z$, $n_e T_{ez}$, and $n_e V_{d,ez}^2$.

The second model features a large library that consists of the relevant plasma state variables and their nonlinear observables, including the product of the linear and the quadratic functions of the plasma variables. For this model, the terms in the library are the following: $n_e$, $V_{d,ez}$, $T_{ex}$, $E_z$, $n_e V_{d,ez}$, $n_e E_z$, $n_e T_{ez}$, $n_e V_{d,ez}^2$, $n_e E_z^2$, $n_e T_{ez}^2$, $n_e^2 V_{d,ez}$, $n_e^2 E_z$, $n_e^2 T_{ez}$, $V_{d,ez} E_z$, $V_{d,ez} T_{ez}$, $E_z T_{ez}$, $V_{d,ez}^2 E_z$, $V_{d,ez} E_z^2$, $V_{d,ez}^2 T_{ez}$, $V_{d,ez} T_{ez}^2$, $E_z^2 T_{ez}$, and $E_z T_{ez}^2$.

Accordingly, the matrices of the coefficients are represented by $\Phi_{SL} \in R^{12 \times 1}$ for the Phi Method model with a small library (SL) and $\Phi_{LL} \in R^{66 \times 1}$ for the model with an expanded large library (LL).

The predicted time evolution signals of the $J_{ex}$ at the midpoint of the domain from the above two Phi Method ROMs are compared against the true data in Figure 12(a) and (b). In addition, Figure 12(c) presents the temporal variation of the loss factors associated with the two models. The loss factors are calculated over the entire domain. It is evident from the plots in Figure 12 that the predictions of the two ROMs are nearly identical.

In Figure 13, the rearranged representations of the $\Phi_{SL}$ and the $\Phi_{LL}$ matrices (operators) are illustrated. Detailed interpretations of the obtained coefficients and their comparison with the analytical values from the discretization of the governing PDEs are discussed in Part II of this article [62].

Here, we would point out from Figure 13 that, even in the presence of numerous dynamical terms in the library, Phi Method can identify the dominant term(s) which, in this special case, is the $n_e T_{ez}$. Also, the relative magnitude of the coefficients of the terms that are shared between the two models are comparable. Nonetheless, for the model with the larger library, the weight of the pressure contribution ($\Pi_{zz}$) is distributed among all observables that have been the functions of both the $n_e$ and the $T_{ez}$. In other words, a portion of the pressure contribution is captured within the $n_e T_{ez}^2$ and the $n_e^2 T_{ez}$ terms, apart from the main share, which is attributed to the $n_e T_{ez}$. As a result, the algorithm assigns non-zero coefficients to the terms $n_e T_{ez}^2$ and $n_e^2 T_{ez}$, even though they are not essential for the representation of the dynamics. This observation is a consequence of the least-square regression ($L2$-norm minimization) corresponding to the current Phi Method's implementation. The $L2$-norm minimization promotes uniformity among the optimized coefficients of the various terms.

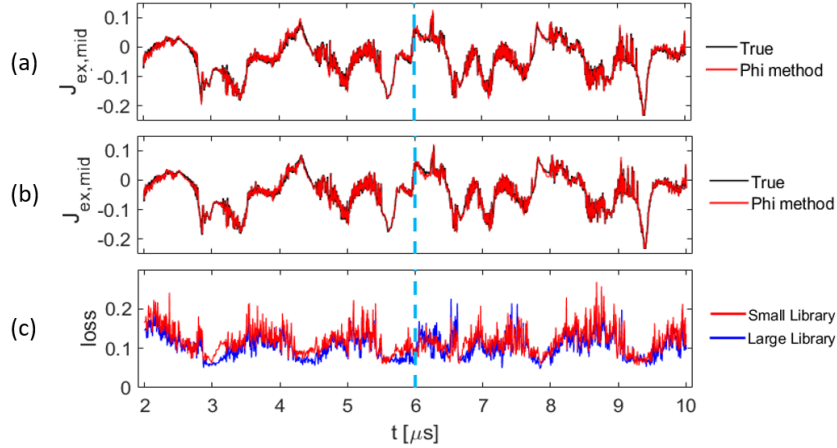

Figure 12: Comparison of the Phi method ROMs for the normalized electrons' axial current density ($J_{ex}$) when regressing over a small and a large library of terms in the test case 3; (a) &(b) time evolutions of the local values of the $J_{ex}$ at the mid-location within the domain from the ground-truth simulation and from the Phi Method models, plot (a) with a small library, plot (b) with a large library of terms. (c) Time evolutions of the loss factor calculated over the entire domain associated with each ROM. Dashed blue lines indicate the end of the training interval.

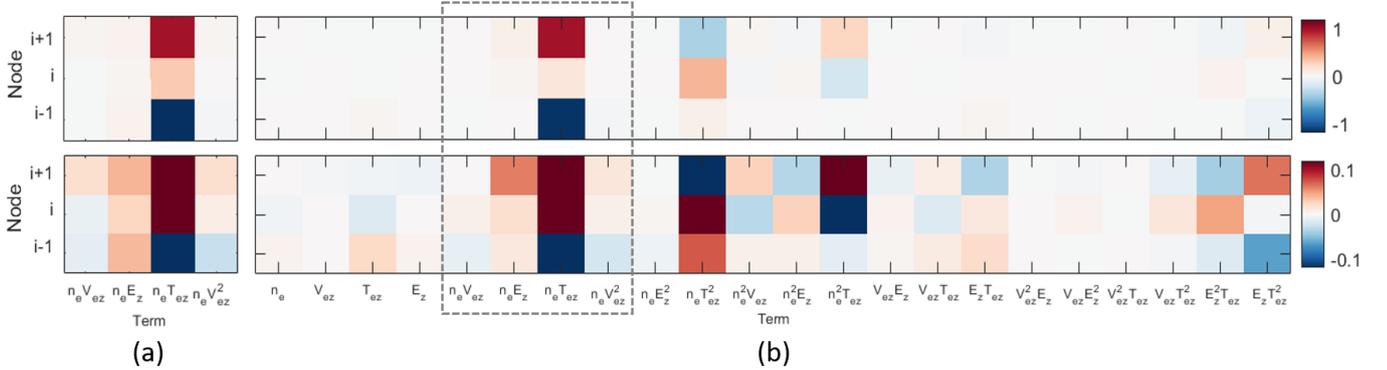

Figure 13: Rearranged normalized representations of the coefficient matrices ($\Phi$) from the Phi method's application to the $J_{ex}$ data of test case 3; (a) $\Phi_{SL}$ visualization (small library of terms), (b) $\Phi_{LL}$ visualization (large library of terms). The bottom row is a rescaled version of the top row. The dashed box encompasses the terms shared between the two Phi Method models.

The uniformity-promoting behavior can be avoided by promoting sparsity in the discovered model, which can be achieved by imposing constraints that encourage the algorithm to incorporate as few terms as required in the model. While the presented case here suggests that both models have equivalent predictive capabilities (from Figure 12), the model with fewer terms is not only more interpretable but also likely to demonstrate better generalizability across a diverse range of scenarios by minimizing the risk of overfitting to the data. Thus, the main purpose of encouraging sparsity is to improve the interpretability and generalizability of the model.

The common approaches to incorporate sparsity in the model include $L1$ regularization (LASSO) and thresholding techniques, such as Sequential Thresholding Least Squares (STLS) [63], and the Sparse Relaxed Regularized Regression (SR3) [63]. In L1 regularization, an additional penalty term is added to the standard loss function of the model, which is proportional to the sum of absolute values of the model's coefficients. This penalty term encourages some of the coefficients to be zero, effectively selecting a subset of the library terms. The thresholding techniques involve eliminating the terms with coefficients below a certain threshold [63]. Hard thresholding sets the coefficients below a pre-defined value to zero, whereas soft thresholding shrinks the coefficients' values toward zero. The incorporation of sparsity-promoting techniques into the Phi Method's algorithm is the subject of future research.

**Section 5: Conclusions**

The aim of this Part I manuscript was to introduce and demonstrate a novel data-driven local operator finding algorithm, termed "Phi Method", that is recently developed at IPPL. The underlying concept and formulation of Phi Method was presented in detail in this paper. It was emphasized that Phi Method is developed to enable the discovery of the discretized form of the differential equations describing a system's dynamics and, thus, to allow straightforward and fast time forecasting of the system state variables. We pointed out that Phi Method belongs to the category of the ML/DD algorithms that identify the local dynamics operators and, hence, discover the differential equation(s) that describe the time evolution (dynamics) of a system. An example of such DD algorithms was mentioned to be SINDy. We highlighted that Phi Method is, in essence, an extension to SINDy algorithm to enable the simultaneous discovery of the dynamics and the optimum discretization stencil for the involved dynamics variables.

We demonstrated the remarkable potentials and generalizability of the Phi Method approach to enable the development of accurate and predictive data-driven ROMs across three test cases: (1) a canonical Lorenz attractor problem, (2) a widely adopted 2D fluid benchmark problem corresponding to the case of viscous flow around a cylinder, and (3) a 1D E × B plasma problem representative of the azimuthal dynamics of a Hall thruster discharge. The involved dynamics in the first test case is governed by an ODE, whereas the evolution of the system in the fluid and the plasma test cases are governed by a system of PDEs.

To have a comparative reference for the predictive performance of the Phi-Method-derived ROMs, we applied OPT-DMD to the data of the test cases 2 and 3 and presented the predictions of the resulting ROMs as well. We also provided an overview of OPT-DMD and its underlying formulation, which allowed us to elaborate on the similarities and the differences between the global OPT-DMD and the local Phi Method techniques toward data-driven reduced-order modelling.

Across all the three test cases, Phi Method was shown to reliably learn the involved dynamics, and the obtained ROMs from this approach featured a great predictive capacity. With regard to the most notable outcomes from all the verifications presented in this paper, in test case 1, we verified the interpretability of the Phi Method's learned operator (the $\Phi$ matrix). We showed that the dominant coefficients of the $\Phi$ matrix had values that compared with a difference of less than 1 % against the analytical coefficients from the discretized ODE that describes the dynamics of the Lorenz attractor.

In test case 2, we showed that a linear Phi Method ROM, which excludes the nonlinear terms that are present in the PDE describing the evolution dynamics of the vorticity field, provides a reasonably well representation of the dynamics and provides predictions that compare closely with the ground-truth data, especially in terms of the time variations of the local vorticity values. We also demonstrated that Phi Method can learn the underlying dynamics from the data of the transient state of the system, and the resulting model also provides accurate predictions. In this sense, Phi Method exceeds the capabilities of DMD-based data-driven techniques, which are inherently limited in their ability to properly represent the transient phenomena. The capability of Phi Method to work seamlessly with the transient data is particularly beneficial for the dynamics discovery of systems that may not exhibit a quasi-steady state or that for which the generation of statistically representative high-fidelity data of the quasi-steady state of the system may be computationally burdensome, such as the case of a Hall thruster discharge.

In test case 3, the ROMs from Phi Method and OPT-DMD were successfully demonstrated toward a simultaneous forecasting of several plasma properties. Phi Method was overall observed to provide a more accurate prediction of the variations in the local values of the plasma state variables compared to OPT-DMD throughout the training and test period, an observation that was linked to the inherently local learning of the dynamics associated with the Phi Method algorithm.

Future work on Phi Method to further mature the approach will involve the incorporation of sparsity-promoting optimization within the algorithm in order to progress toward fully self-consistent learning of the closed set of PDEs describing the evolution of the plasma state from high-fidelity data.

From a broader applied perspective, we would emphasize that the negligible computational cost of the DD ROMs from Phi Method and OPT-DMD compared to the kinetic PIC simulations in addition to the reasonable predictive performance of the ROMs from these methods as was demonstrated in this work make them highly suitable as the foundational building blocks of the digital twins for the plasma propulsion systems. Indeed, for practical digital twins that can enable computer-aided design and qualification of next-generation plasma thrusters, a fast, robust, and reliable prediction and forecasting of the systems' dynamics is key.

**Acknowledgments**:

The present research is carried out within the framework of the project "Advanced Space Propulsion for Innovative Realization of space Exploration (ASPIRE)". ASPIRE has received funding from the European Union's Horizon 2020 Research and Innovation Programme under the Grant Agreement No. 101004366. The views expressed herein can in no way be taken as to reflect an official opinion of the Commission of the European Union.

MR, FF, and AK gratefully acknowledge the computational resources and support provided by the Imperial College Research Computing Service (http://doi.org/10.14469/hpc/2232).

**Data Availability Statement**:

The simulation data that support the findings of this study are available from the corresponding author upon reasonable request.

**Appendix: Assessment of the effects of noise in the data**

The effects of noise in the data on the performance of the ROMs from Phi Method are investigated in this appendix for the test cases 2 and 3. The noise robustness of Phi Method is assessed comparatively against that of OPT-DMD, which was demonstrated in Ref. [13] to exhibit strong noise robustness characteristics.

To perform the noise robustness analyses, artificial Gaussian noise was added to the simulation data of the test cases 2 and 3, and the data-driven models were trained on the resulting noisy data.

**A. Test case 2**

For this test case, we added Gaussian noise with the standard deviation of $\sigma = 0.03$ to all the flow properties. The characteristics of the Phi Method and the OPT-DMD ROMs used here are kept the same as those that were described in subsection 3.2.1.

Figure 14 and Figure 15 present the predictions of the ROMs trained on the noisy data. The predictions are shown in terms of the time evolutions of the spatially averaged vorticity and the local value at the mid-domain-location of the vorticity field in Figure 14. Figure 15 presents the predicted snapshots of the entire vorticity field at a few sample time instants within the test window. The true data in these figures represents the simulation data after the introduction of noise. In Figure 16, the absolute-difference plots between the true and the predicted vorticity field snapshots are additionally illustrated.

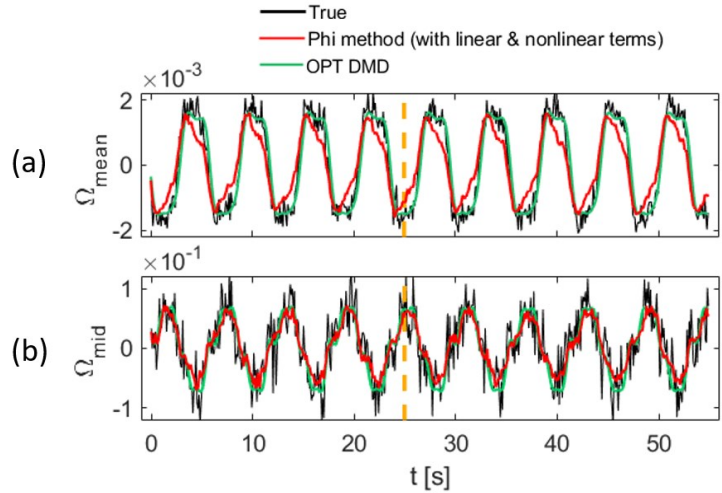

Figure 14: Comparison between the predictions of the Phi Method and the OPT-DMD models with respect to the ground-truth data in the presence of a Gaussian white noise in the test case 2; time evolutions of (a) spatially averaged normalized vorticity, (b) local value of the vorticity at the mid-location within the simulation domain. Yellow lines indicate the end of the training interval.

These results show that both the Phi method and the OPT-DMD ROMs can effectively learn the underlying dynamics in the presence of noise. This is not a surprising outcome for OPT-DMD because the noise is largely excluded after performing a low-rank ($r = 14$) truncation of the SVD/POD expansion of the data as part of the DMD algorithm. However, it is a notable observation that Phi Method is also able to maintain its predictive capability without featuring a significant noise level in its predictions.

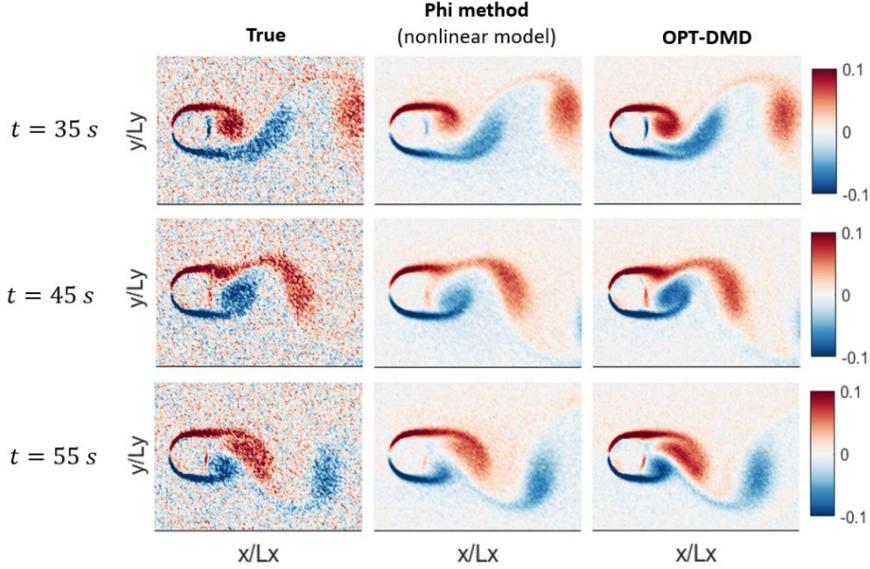

Figure 15: Comparison between the predictions of the Phi Method and the OPT-DMD models with respect to the ground-truth data in the presence of a Gaussian white noise in the test case 2; snapshots of the normalized vorticity field at three different time instants during the test interval.

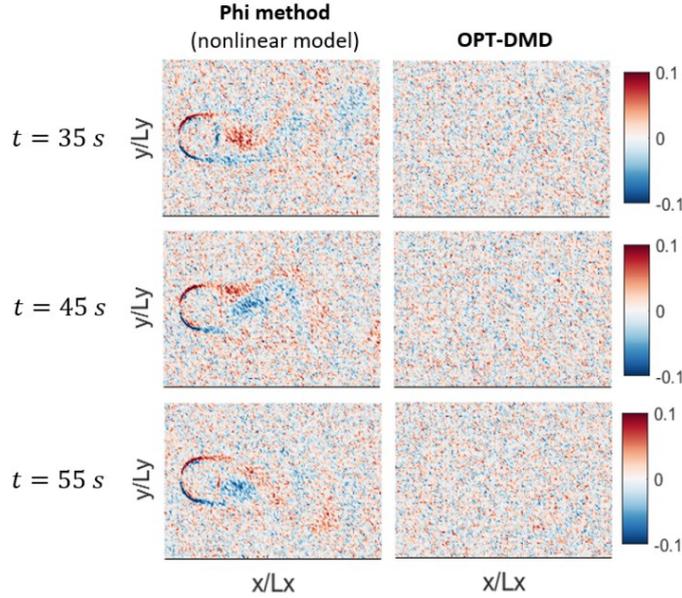

Figure 16: Comparison between the Phi Method and the OPT-DMD models in the presence of a Gaussian white noise in the test case 2; plots of absolute difference between the true and the predicted normalized vorticity snapshots at three different time instants during the test interval.

## B. Test case 3

Gaussian noise with a standard deviation of $\sigma = 0.05$ is added to the full dataset corresponding to all the relevant plasma properties obtained from the PIC simulation of the 1D azimuthal plasma discharge case. It is emphasized that the original PIC data inherently contains a certain level of (statistical) noise, and the Gaussian noise is added on top of that pre-existing noise. As a result, this case represents a more extreme scenario than the one analyzed in the previous appendix section, noting also that the introduced noise has a higher amplitude.

The Phi Method and the OPT-DMD models, whose details were presented in subsection 3.3.1, were trained on the dataset with augmented noise. The time variations of the predicted values of the plasma properties at the midpoint of the domain, and the complete spatiotemporal maps of the plasma properties from the Phi Method and the OPT-DMD ROMs are provided in Figure 17 and Figure 18, respectively. We observe that OPT-DMD effectively filters the noise and is able to provide a relatively consistent representation of the system. For the case of Phi Method, however, the presence of high-level noise in the data affects the obtained ROM, resulting in the prediction of lower-amplitude fluctuations in the plasma properties' signals compared to the ground-truth. This is because certain dynamical information within the data is obscured by the noise. Nevertheless, apart from the amplitude, the main traces of the signals and the frequencies of the fluctuations across the different properties are consistently captured by the Phi Method ROM.

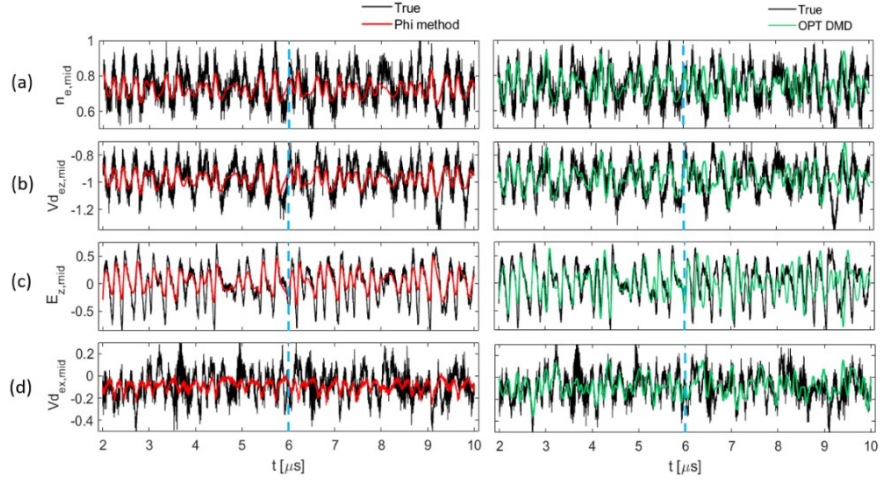

Figure 17: Comparison of the predictions from the Phi Method (left column) and the OPT-DMD (right column) ROMs against the ground-truth data in the presence of a Gaussian white noise in the test case 3; time evolution of the local values at the mid-location within the simulation domain of (a) normalized electron number density ($n_e$), (b) normalized electrons' azimuthal drift velocity ($V_{d,ez}$), (c) normalized azimuthal electric field ($E_z$), and (d) normalized electrons' axial drift velocity ($V_{d,ex}$). Dashed blue lines indicate the end of the training interval.

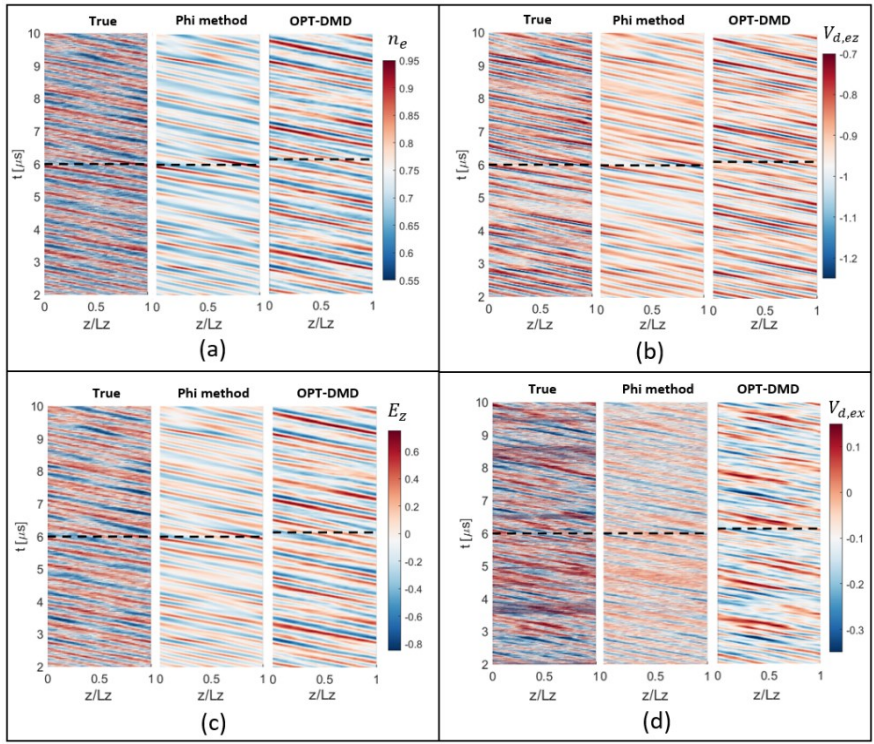

Figure 18: Comparison of the predictions from the Phi Method and the OPT-DMD ROMs against the ground-truth data in the presence of a Gaussian white noise in the test case 3; spatiotemporal evolution plots of (a) normalized electron number density ($n_e$), (b) normalized electrons' azimuthal drift velocity ($V_{d,ez}$), (c) normalized azimuthal electric field ($E_z$), and (d) normalized electrons' axial drift velocity ($V_{d,ex}$). Dashed black lines indicate the end of the training interval.